\documentclass{article}
\usepackage{arxiv}
\usepackage[utf8]{inputenc}
\usepackage[T1]{fontenc}
\usepackage{xcolor}
\usepackage[sort&compress, round]{natbib}
\definecolor{blendedblue}{rgb}{0.2, 0.2, 0.6}
\usepackage{graphicx}
\usepackage{enumerate}
\usepackage[bookmarks   = true,
            citecolor   = blendedblue,
            colorlinks  = true,
            linkcolor   = blendedblue,
            urlcolor    = blendedblue,
            citecolor   = blendedblue,
            linktocpage = false,
            hyperindex  = true]{hyperref}
\usepackage{booktabs}
\usepackage{amsfonts}
\usepackage{nicefrac}
\usepackage{amsmath, amssymb, amsthm}
\usepackage{url}
\usepackage{doi}
\usepackage{bm}
\usepackage{todonotes}
\usepackage{fontawesome}
\usepackage{mathtools}
\usepackage{lmodern}
\usepackage{siunitx}
\usepackage{booktabs}
\usepackage{etoolbox}
\usepackage{calc}
\usepackage{enumitem}
\usepackage{dsfont}
\usepackage{multirow}
\usepackage{array}
\usepackage{graphicx}
\definecolor{blendedblue}{rgb}{0.2, 0.2, 0.6}

\newcommand{\E}{{\mathbb{E}}}
\newcommand{\Cov}{{\operatorname{Cov}}}

\usepackage{enumerate} 
\usepackage{orcidlink}

\usepackage[labelsep=period]{caption}

\renewcommand{\headeright}{}

\usepackage{parskip}
\makeatletter 
\renewcommand\@biblabel[1]{#1.}
\makeatother

\title{Covariate-Adjusted Functional Data Analysis for Structural Health Monitoring}
\renewcommand{\shorttitle}{CAFDA-SHM: Covariate-Adjusted Functional Data Analysis for Structural Health Monitoring}
\date{}

\author{Philipp Wittenberg\orcidlink{0000-0001-7151-8243}\\
 	Dept. of Mathematics and Statistics\\
	School of Economics and Social Sciences\\
        Helmut Schmidt University\\
	Hamburg, Germany\\
	\texttt{pwitten@hsu-hh.de} \\
	\And
        Lizzie Neumann\orcidlink{0000-0003-2256-1127}\\
 	Dept. of Mathematics and Statistics\\
	School of Economics and Social Sciences\\
        Helmut Schmidt University\\
	Hamburg, Germany\\
	\texttt{neumannl@hsu-hh.de} \\
	\And
        Alexander Mendler\orcidlink{0000-0002-7492-6194}\\
        Dept. of Materials Engineering\\
        TUM School of Engineering and Design\\
	Technical University of Munich\\
        Munich, Germany\\
	\texttt{alexander.mendler@tum.de}\\
	\And 
        Jan Gertheiss\orcidlink{0000-0001-6777-4746}\\
 	Dept. of Mathematics and Statistics\\
 	School of Economics and Social Sciences\\
        Helmut Schmidt University\\
	Hamburg, Germany\\
	\texttt{gertheij@hsu-hh.de} \\
}

\begin{document}	
%%%%%%%%%%%%%%%%%%%%%%%%%%%%%%%%%%%%%%%%%%%%%%%%%%%%%%%%%
\maketitle

\begin{abstract}
Structural Health Monitoring (SHM) is increasingly applied in civil engineering. One of its primary purposes is detecting and assessing changes in structure conditions to increase safety and reduce potential maintenance downtime. Recent advancements, especially in sensor technology, facilitate data measurements, collection, and process automation, leading to large data streams. We propose a function-on-function regression framework for (nonlinear) modeling the sensor data and adjusting for covariate-induced variation. Our approach is particularly suited for long-term monitoring when several months or years of training data are available. It combines highly flexible yet interpretable semi-parametric modeling with functional principal component analysis and uses the corresponding out-of-sample Phase-II scores for monitoring. The method proposed can also be described as a combination of an ``input-output'' and an ``output-only'' method.
\end{abstract}
\vfill

%%%%%%%%%%%%%%%%%%%%%%%%%%%%%%%%%%%%%%%
\section{Introduction}
%%%%%%%%%%%%%%%%%%%%%%%%%%%%%%%%%%%%%%%
Structural health monitoring (SHM) employs sensor technologies to collect data from structures such as bridges to detect, localize, or quantify damage \citep{Deraemaeker.etal_2008, Kullaa_2011, Hu.etal_2012, Magalhaes.etal_2012}. These field measurements often exhibit missing data and are influenced by environmental and operational factors such as temperature, wind, humidity, or traffic load \citep{Wang.etal_2022}. Multiple studies, including \cite{Wang.etal_2022, Han.etal_2021} and the references therein, identify temperature as a predominant factor affecting structural stiffness and material properties due to thermal expansion and contraction \citep{Han.etal_2021}. Consequently, data-driven methods are essential to take these confounding effects in SHM data analysis into account. 

With respect to separating temperature-induced responses from structural responses, it can be distinguished between so-called ``input-output'' methods and ``output-only'' approaches~\citep{Wang.etal_2022}. In the first case, both the sensor measurements and observations of the confounding variables, such as temperature, are considered, while in the latter case, as the name suggests, only the responses of the structures are used, often using projection methods such as principal component analysis (PCA). Among input-output methods, a prevalent approach is regressing sensor measurements on environmental or operational variables, also known under the name \emph{response surface} modeling. Then, following the so-called ``subtraction method'', the predicted sensor data is subtracted from the actually observed data, and the residuals are used for further analysis. For fitting regression function(s) to the data, various methods have been used in the SHM literature, ranging from simple linear or polynomial regression to advanced machine learning approaches such as artificial neural networks, see \cite{Avci.etal_2021}. 
However, often, these methods ignore that output data may exhibit daily or yearly patterns or that error terms are correlated over time, e.g., if estimating unknown parameters through least-squares \citep{Cross.etal_2013, Maes.etal_2022}. In other cases, rather restrictive parametric assumptions known from time series analysis are made \citep{Hou.Xia_2021}. Furthermore, input-output and output-only methods are typically considered to be completely different approaches. An overarching framework is still missing where, for instance, it is possible to switch from one approach to the other without changing the downstream monitoring procedure, or the common situation can be handled where measurements are available only for a few confounding variables while others are unobserved or unknown.

In light of this, we will present a novel functional data analysis (FDA) perspective to address these challenges in SHM. FDA has been an area of intensive methodological research over the last two or three decades. For an introduction to FDA in general and an overview of recent developments, the works by \cite{Ramsay.Silvermann_2005, Wang.etal_2016}, and \cite{Gertheiss.etal_2024} are recommended. Existing applications of FDA within SHM are reviewed by \cite{Momeni.Ebrahimkhanlou_2022}. In SHM, so far, FDA has primarily been used for distributional regression and change point detection. For instance, \cite{Chen.etal_2020} used FDA to segment data by time and analyze corresponding probability density functions through warping functions and functional principal component analysis (FPCA). Other notable contributions include the works by \cite{Chen.etal_2018}, \cite{Chen.etal_2019}, \cite{Chen.etal_2021b}, who developed methods for imputing missing data using distribution-to-distribution and distribution-to-warping functional regression, and \cite{Lei.etal_2023b}, \cite{Lei.etal_2023}, \cite{Lei.etal_2023c} focused on outlier detection and change-point detection in FDA. \cite{Jiang.etal_2021} modeled temperature-induced strain relationships using warping functions and FPCA.

\begin{figure}[bt]
\centering
\includegraphics[scale=.5]{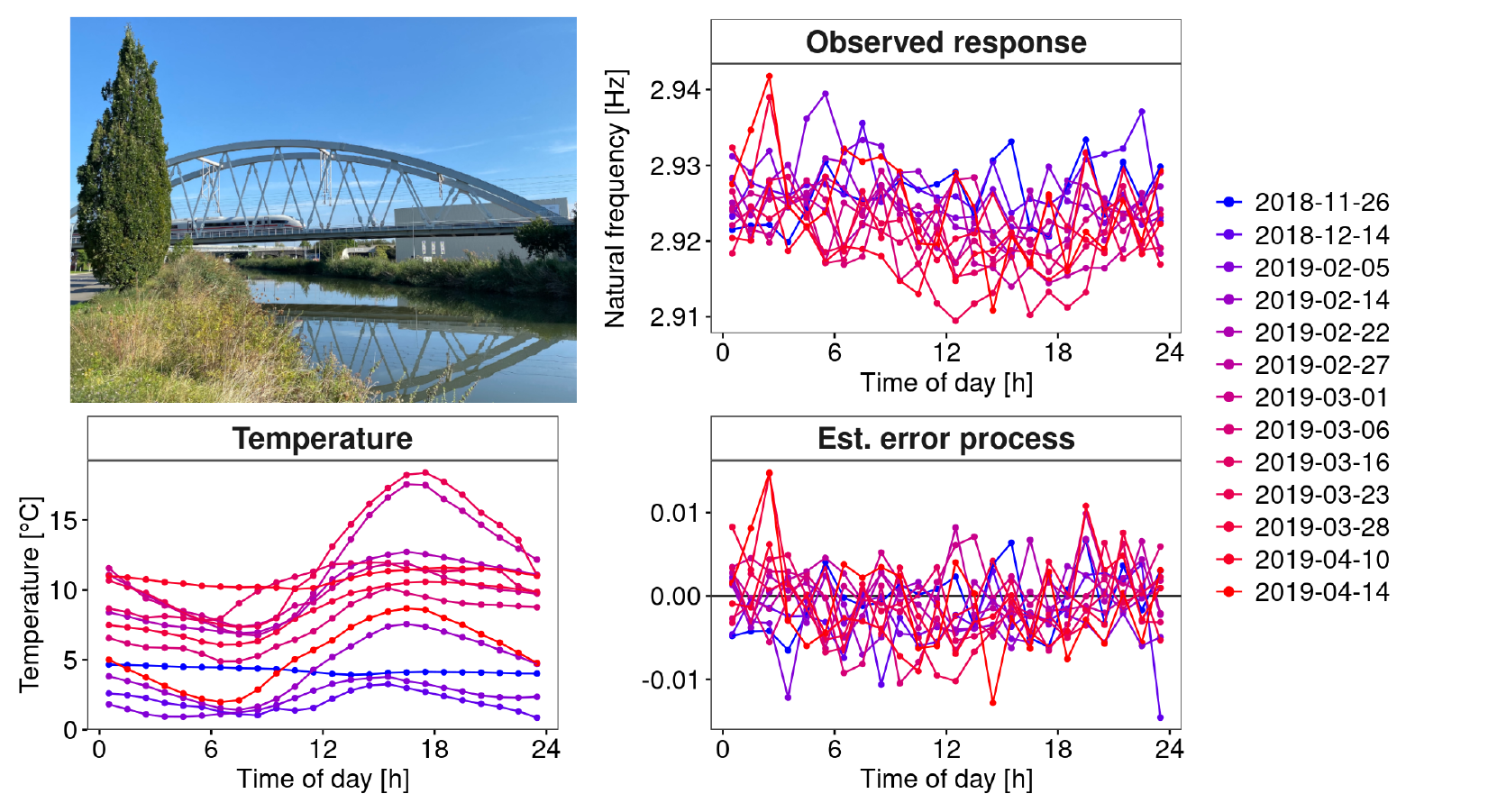}
\caption{KW51 Bridge from the south side in September 2024 (top-left). The natural frequency of mode 6 for some selected days before the retrofitting started (top-right) and the corresponding steel temperature curves (bottom-left). If a simple piecewise linear model is used for temperature adjustment, the error profiles (bottom-right) are obtained.}
\label{fig:KW51_bilinear}
\end{figure}

To elucidate our functional perspective, consider Figure~\ref{fig:KW51_bilinear}. Similar plots in SHM studies, such as \cite{Xia.etal_2012}, \cite{Xia.etal_2017} and \cite{Zhou.etal_2011}, have been used to illustrate the relationship between natural frequencies, strain or displacement, and temperature, but without exploiting the functional nature of the data. The data in the top-left panel of Figure~\ref{fig:KW51_bilinear} is from the KW51 railway bridge~\citep{Maes.Lombaert_2020, Maes.Lombaert_2021}, near Leuven, Belgium, monitored from October 2nd, 2018, to January 15th, 2020, including a retrofitting period from May 15th to September 27th, 2019. This dataset contains hourly natural frequencies and steel surface temperature (more details on the bridge will be given in Section \ref{sec:Results}). The top-right panel of Figure~\ref{fig:KW51_bilinear} shows the bridge's natural frequency (mode 6) for selected days before retrofitting. It appears plausible to assume that there is some underlying daily pattern common to all profiles, which might be caused by environmental influences that may follow some recurring pattern as well. Those patterns and relationships can be further analyzed through FDA, among other things, by regressing the natural frequency profiles on the temperature curves shown in the bottom-left panel of Figure~\ref{fig:KW51_bilinear}. From the colors, it is already visible that there is a negative association between temperature and the natural frequency, meaning that lower temperatures lead to a higher frequency; compare, e.g., \cite{Xia.etal_2006}, \cite{Xia.etal_2012}. The bottom-right panel shows the resulting error profiles when using a conventional, non-functional piecewise linear model to account for this effect as done by \cite{Maes.etal_2022}.

If considering entire daily profiles rather than individual measurements as the sampling instances, this regression task is called ``function-on-function'' regression. In the statistical process monitoring (SPM) literature, it is part of so-called profile monitoring, see \cite{Woodall.etal_2004, Woodall_2007, Maleki.etal_2018} for detailed reviews on that topic. Current state-of-the-art methodologies, such as the (linear) functional regression control chart (FRCC) by \cite{Centofanti.etal_2021} have limitations in SHM applications, where the relationship between temperature and natural frequencies is often nonlinear \citep{Han.etal_2021}. Additionally, by integrating over the entire domain (i.e., the entire day in Figure~\ref{fig:KW51_bilinear}), the employed functional linear model allows ``future observations'' to influence the current value of the response~\citep{Wittenberg.etal_2024}. Specifically, if (material) temperature data is available directly from the structure itself instead of ambient temperature, it appears more plausible to use a so-called concurrent model, where the response data at time point $t$ depends only on the steel temperature at time $t$, not the temperature over the entire day. Finally, the assumption of complete datasets, e.g., made in~\cite{Capezza.etal_2023c}, is impractical because SHM sensor data often exhibit dropouts due to power or sensor failure. To address these challenges often found in SHM, this paper will consider the very general framework of functional additive mixed models~\citep{Scheipl.etal_2015, Greven.Scheipl_2017} instead of common functional linear models used by \cite{Centofanti.etal_2021} and \cite{Capezza.etal_2023c}. Besides the regression task, the functional perspective on SHM data offers an elegant way of ``de-noising'' the data and extracting features through functional principal component scores that can be used as inputs to advanced multivariate control charts. The methods presented in this article can be applied directly to sensor measurements (e.g., strains, inclinations, accelerations) or extracted damage-sensitive features (e.g., eigenfrequencies). That is why this paper will use the more general term ``system outputs''. In summary, using the methods discussed, we can model (i) recurring daily and yearly patterns as well as (ii) confounding effects flexibly and interpretably. Furthermore, the presented functional approach also accounts for (iii) correlations and (iv) heteroscedasticity within error profiles. Finally, it will (v) extract covariate-adjusted features from the system outputs and (vi) unify all these aspects within a coherent statistical framework called \emph{Covariate-Adjusted Functional Data Analysis for SHM} (CAFDA-SHM). This combined approach includes subtraction methods, PCA-based projections (both often used in SHM), and the FRCC in one universal framework, encompassing aspects that have only been considered partially in previous approaches. 

The remainder of the article is structured as follows. In Section \ref{sec:FDA}, our functional data approach is introduced, and our model training and monitoring workflow is presented. The results of a Monte Carlo simulation study validating the developed methods is presented in Section \ref{sec:Simulation}.
A structural health monitoring application for dynamic response data of the KW51 railway bridge and a second case study for a quasi-static response of a reinforced concrete bridge can be found in Section~\ref{sec:Results}, respectively. Section \ref{sec:Conclusion} offers some concluding remarks. Appendix~\ref{sec:model_extensions} provides some further modeling options beyond Section~\ref{sec:FDA}. 
%%%%%%%%%%%%%%%%%%%%%%%%%%%%%%%%%%%%%%%%%%%%%%%%%%%%%%%%%%%%%%%%%%%%%%%%%%%%%%%%%%%%%%%%%%%%%%%%%
\section{A functional data approach for modeling and monitoring system outputs}\label{sec:FDA}
%%%%%%%%%%%%%%%%%%%%%%%%%%%%%%%%%%%%%%%%%%%%%%%%%%%%%%%%%%%%%%%%%%%%%%%%%%%%%%%%%%%%%%%%%%%%%%%%%
%%%%%%%%%%%%%%%%%%%%%%%%%%%%%%%%%%%%%%%%%%%%%%%%
\subsection{Model structure}\label{subsec:model}
%%%%%%%%%%%%%%%%%%%%%%%%%%%%%%%%%%%%%%%%%%%%%%%%

\subsubsection{Basic model}

The model we assume for ``in-control'' (IC) data has the following basic form. To keep things simple, we first restrict ourselves to a single, functional covariate $z_j(t)$, e.g., denoting the temperature at time $t \in \mathcal{T}, \mathcal{T} = (0\text{h}, 24\text{h}]$, and day $j$, and a single system output $u_j(t)$. The latter could be a rather raw sensor measurement (yet preprocessed in some sense), such as strain or inclination data, or extracted features, such as natural frequencies. Then, we assume the basic model
\begin{equation}\label{eq:basic_model}
 u_j(t) = \alpha(t) + f(z_j(t)) + E_j(t), 
\end{equation}
where $\alpha(t)$ is a fixed functional intercept, $f(z_j(t))$ is a fixed, potentially non-linear effect of temperature, and $E_j(t)$ is a day-specific, functional error term with zero mean and a common covariance, i.e., $\E(E_j(t)) = 0$, $\Cov(E_j(s), E_j(t)) = \Sigma(s,t)$, $s,t \in \mathcal{T}$. In the FDA framework used here, sampling instances are days instead of single measurement points, and the daily profiles are considered the quantities of interest. This model has two advantages over scalar-on-scalar(s) regression as typically used for response surface modeling in SHM:
\begin{itemize}
\item The functional intercept $\alpha(t)$ captures recurring daily patterns that cannot be explained through the available environmental or operational variables, e.g., because the factors causing them are not recorded/available. In the case of long-term monitoring, we may extend the one-dimensional $\alpha(t)$ to a two-dimensional surface $\alpha(t,d_j)$, where $d_j$ denotes the time and date of the year corresponding to day $j$ in the data set. The latter accounts for a potential second, i.e., yearly level of periodicity.
\item The error term $E_j(t)$ is typically not a white noise process but correlated over time, i.e., in the $t$-direction. Furthermore, variances may vary over the day. For instance, error variances may be lower at night (e.g., because there is less traffic, no influence by the sun, etc.). In other words, $\Sigma(s,t)$ is not necessarily zero for $s \neq t$, and $\Sigma(t,t)$ is not constant. For illustration, the (estimated) error process for some selected days for the KW51 data is shown in Figure~\ref{fig:KW51_bilinear}~(right), where a piecewise linear model with one breakpoint was used for temperature adjustment, as suggested in the literature \citep{Moser.Moaveni_2011, Worden.Cross_2018, Maes.etal_2022}. Apparently, those profiles exhibit some more structure than pure white noise. However, ignoring this correlation when fitting $\alpha$ and $f$ through ordinary least squares or common maximum likelihood assuming conditional independence between measurements will typically lead to less accurate estimates. More importantly, if (conditional) independence is assumed but not given, measures of statistical uncertainty, such as confidence and prediction intervals or control limits, will be biased. In SHM, this can be particularly harmful as these quantities are used to determine if measurements are ``out-of-control''. For instance, if a memory-based control chart with the assumption of independence is used to detect a mean shift in an error process as shown in Figure~\ref{fig:KW51_bilinear}, the false positive rate will be substantially increased \citep{Knoth.Schmid_2004}. The big advantage of the FDA framework over standard parametric assumptions known from time series analysis, such as auto-correlation of some specific order, is that the error process can be modeled in a very flexible, semi-parametric fashion through functional principal component analysis.  
\end{itemize}
In the case of SHM, there is another important aspect to consider with respect to $E_j(t)$: This process contains the relevant information for the monitoring task since it captures deviations from the system output $\alpha(t) + f(z_j(t))$ that would be expected for a specific, let us say, temperature at time $t$ if the structure is ``in-control''. For exploiting this information, it is beneficial to further decompose this process into a more structural component $w_j(t)$ and white noise $\epsilon_j(t)$ with variance $\sigma^2$, i.e.,
\begin{equation}\label{eq:Ej1}
E_j(t) = w_j(t) + \epsilon_j(t).
\end{equation}
Since $\epsilon_j(t)$ is assumed to be pure noise, it does not carry relevant information, and $w_j(t)$ should be the part to focus on for monitoring purposes. To decompose $E_j(t)$ into $w_j(t)$ and $\epsilon_j(t)$, we use functional principal component analysis (FPCA). It is based on the Karhunen-Loeve expansion \citep{Karhunen_1947,Loeve_1946}, which allows us to expand the random function $E_j(t)$ as
\begin{equation}\label{eq:Ej2}
E_j(t) = \sum_{r=1}^\infty \xi_{rj}\phi_r(t),    
\end{equation}
where $\phi_r$ are orthonormal eigenfunctions of the covariance, i.e., $\int_{\mathcal{T}} \phi_r(t) \phi_k(t) dt = 1$ if and only if $k=r$ and zero otherwise. In particular, according to Mercer's theorem \citep{Mercer_1909},
\begin{equation}\label{eq:mercer}
\Cov(E_j(s), E_j(t)) = \Sigma(s,t) =
\sum_{r=1}^\infty \nu_r \phi_r(s)\phi_r(t)
\end{equation}
with decreasing eigenvalues $\nu_1 \geq \nu_2 \geq \dots \geq 0$. Furthermore, $\xi_{rj}$ are uncorrelated random scores with mean zero and variance $\nu_r$, $r = 1, 2, \dots$, and are independently normal if $E_j$ is a Gaussian process.

In FPCA, the sum in \eqref{eq:Ej2} is truncated at a finite upper limit $m$, which gives the best approximation of $E_j$ with $m$ basis functions~\citep{Rice.Silvermann_1991}. Looking at the fraction  $\{\sum_{r=1}^m \nu_r\} / \{\sum_{r=1}^{\infty} \nu_r\}$ allows a quantitative assessment of the variance explained by the approximation, and the concrete value of $m$ is typically chosen such that 95\% or 99\% of the overall variance is explained. By choosing such a large value, it is reasonable to assume that the selected $m$ eigenfunctions account for all relevant features in the data and the remainder is merely noise. Consequently, we set   
\begin{equation}\label{eq:wj}
w_j(t) = \sum_{r=1}^m \xi_{rj}\phi_r(t),
\end{equation}
and use the scores $\xi_{1j},\ldots,\xi_{mj}$ as damage-sensitive features for monitoring. As the functions $\alpha, f, \phi_1, \ldots, \phi_m$ are estimated from IC data in the model training phase, it is the scores obtained for future data that tell us whether the system outputs deviate from the values that would be expected for an IC structure over the day for given values of the covariate (and the specific time of the year, if $\alpha(t,d_j)$ instead of $\alpha(t)$ is used in model~\eqref{eq:basic_model}). How to estimate the different model components, such as $\alpha, f, \phi_1, \ldots, \phi_m$ from the training data, and how to estimate the scores on future data, will be described in Section~\ref{subsec:fitting} below. At this point, it is important to note that if environmental and operational variables are present beyond $z$, and those variables are measured, model~\eqref{eq:basic_model} can be extended to include multivariate covariates, see Section \ref{subsec:mod_and_ext}. If there are additional confounding effects that are not measured or not known at all, we can proceed analogously to the popular output-only method using (multivariate) PCA. 
That means we assume that the first, let us say, $\varrho$ components mainly account for variation induced by latent factors \citep{Cross.etal_2012} so we only use the remaining scores $\xi_{jr}$, $r > \varrho$, for monitoring. An important advantage of model~\eqref{eq:basic_model} over output-only methods is that available covariate information can still be exploited through $f$. How to modify our approach in terms of a pure output-only method if no covariate information is available at all will be described in the next paragraph.

\subsubsection{Modifications and extensions}\label{subsec:mod_and_ext}

Our basic model \eqref{eq:basic_model} is a particular case of a functional additive mixed model, as introduced and discussed by \cite{Scheipl.etal_2015}, \cite{Scheipl.etal_2016}, and \cite{Greven.Scheipl_2017}, which can be simplified, modified, or extended in various ways depending on the available data and prior knowledge. Some alternative specifications that may be an option with the data available in our case studies in Section~\ref{sec:Results} are recapped as follows. 
\begin{itemize}
    \item \textit{Standard linear models:} If we have reason to believe that potential daily patterns can be explained entirely through variations in $z$, we can replace $\alpha(t)$ with a constant $\alpha_0$. Similarly, if it is believed that the association of the confounder $z$ and system output $u$ is linear, we can replace $f(z_j(t))$ in~\eqref{eq:basic_model} with $\beta z_j(t)$. So, the standard linear regression model is a special case in our broader framework. Typically, if fitting the more flexible model~\eqref{eq:basic_model} to the data, a nearly constant $\alpha$ and a close to linear $f$ will indicate that the simpler model is sufficient. 
    \item \textit{Unmeasured covariates:} If covariate information is unavailable, $f(z_j(t))$ can be omitted, and our approach turns into an output-only method using FPCA. In that case, replacing $\alpha(t)$ with $\alpha(t,d_j)$ is highly recommended. The reason for this is that variables such as temperature typically vary over the day and year, imposing a specific yearly pattern on $u$ as well. The latter can be modeled through $\alpha(t,d_j)$. 
    \item \textit{Multiple covariates:} If $p$ covariates $z_{j1}(t),\ldots,z_{jp}(t)$, e.g., temperature, relative humidity, wind speed, solar radiation, have an effect, or if several temperature sensors are used to account for the local temperatures at different spatial locations in the construction material, their effects can be combined additively in~\eqref{eq:basic_model} by replacing the term $f(z_j(t))$ with $\sum_{k=1}^p f_k(z_{jk}(t))$.
    \item \textit{Covariate interactions:} If $z_{jk}(t), z_{jl}(t)$ are believed to have an interacting effect on the system output, e.g., temperature and relative humidity, model~\eqref{eq:basic_model} can also contain (two-way) interactions in terms of $f_{kl}(z_{jk}(t),z_{jl}(t))$. In theory, interactions of higher order are possible as well. However, it can be challenging to estimate the corresponding parameter functions due to the computing resources and amount of data needed to learn those interactions reliably, compare Section~\ref{subsec:fitting}.  
\end{itemize}
While all those models are so-called \emph{concurrent} models, which are useful to adjust for the temperature of the structure itself, the framework of functional additive mixed models is flexible enough to cope with delayed effects (as, e.g., plausible for ambient temperature)  using so-called \emph{historical} functional effects \citep{Scheipl.etal_2016}. Also, the very popular linear function-on-function regression approach \citep{Centofanti.etal_2021} is included as a special case. An overview of various modeling options beyond those discussed above is provided in Appendix~\ref{sec:model_extensions}.

%%%%%%%%%%%%%%%%%%%%%%%%%%%%%%%%%%%%%%%%%%%%%%%%%%%
\subsection{Model training strategy}\label{subsec:fitting}
%%%%%%%%%%%%%%%%%%%%%%%%%%%%%%%%%%%%%%%%%%%%%%%%%%%
Model fitting is carried out on in-control/Phase-I training data only. This will be described in Sections~\ref{subsubsec:basic} to \ref{subsubsec:fpca}. 
In Section~\ref{subsubsec:scores}, we will discuss how to obtain the scores for Phase-II data that can be used for monitoring.
%%%%%%%%%%%%%%%%%%%%%%%%%%%%%%%%%%%%%%%%%%%%%%%%%%%%%%%%%%%
\subsubsection{Basic model training}\label{subsubsec:basic}
%%%%%%%%%%%%%%%%%%%%%%%%%%%%%%%%%%%%%%%%%%%%%%%%%%%%%%%%%%%
For estimating functions such as $\alpha$ or $f$ in \eqref{eq:basic_model}, we follow an approach popular in functional data analysis and semiparametric regression, compare \cite{Greven.Scheipl_2017} and \cite{Wood_2017}. The unknown function, say $f$, is expanded in basis functions such that
\begin{equation}\label{eq:f_basis}
    f(z) = \sum_{l=1}^L \gamma_lb_l(z).
\end{equation}
A popular choice for $b_1(z),\ldots,b_L(z)$ is a cubic B-spline basis, which means that $f$ is a cubic spline function \citep{deBoor_1978, Dierckx_1993}. For being sufficiently flexible with respect to the types of functions that can be fitted through \eqref{eq:f_basis}, typically, a rich basis with a large $L$ is chosen. A large $L$, however, often leads to wiggly estimated functions if the basis coefficients $\gamma_1,\ldots, \gamma_L$ are fit without any smoothness constraint. The latter is typically imposed by adding a so-called \emph{penalty} term when fitting the unknown coefficients through least-squares or maximum likelihood. A popular penalty is the integrated squared second derivative $\int_\mathcal{D} [f''(z)]^2 dz$, where $\mathcal{D}$ is the domain of $f$. An alternative is a (quadratic) penalty on the discrete second or third-order differences of the basis coefficients $\gamma_l$, which gives a so-called P-spline, see \cite{Eilers.Marx_1996} for details. If we assume that the eigenfunctions $\phi_1,\ldots,\phi_m$ from \eqref{eq:wj} are known, model \eqref{eq:basic_model} becomes
\begin{equation}\label{eq:basic_model2}
    u_j(t) = \sum_{l=1}^{L^{(\alpha)}} \gamma^{(\alpha)}_l b^{(\alpha)}_l(t) + \sum_{l=1}^{L^{(f)}} \gamma^{(f)}_l b^{(f)}_l(z_j(t)) + \sum_{r=1}^m \xi_{rj}\phi(t) + \epsilon_j(t),
\end{equation}
where $\gamma^{(\alpha)}_l$, $b^{(\alpha)}_l$, $\gamma^{(f)}_l$, $b^{(f)}_l$ are the basis coefficients and functions for $\alpha$ and $f$, respectively, and $\xi_{rj}$ are day-specific random effects. Given a training data set of system outputs and covariate information $\{u_j(t_{ji}), z_j(t_{ji})\}$ for days $j=1,\ldots, J$ and time points $t_{ji} \in \mathcal{T}$, $i=1,\ldots, N_j$, the unknown parameters $\gamma^{(\alpha)}_l, \gamma^{(f)}_l$ can be estimated through penalized least-squares, whereas the random effects $\xi_{rj}$ can be predicted using (linear) mixed models approaches. In fact, the quadratic penalties mentioned above can be incorporated into the mixed model framework, such that the strength of the penalty can be determined through (restricted) maximum likelihood, analogously to the variance components $\nu_1,\ldots,\nu_m$ and $\sigma^2$, see \cite{Wood_2011, Wood_2017} for details. An in-depth discussion of these methods is beyond the scope of this paper. However, an important point to note is that model \eqref{eq:basic_model} and hence \eqref{eq:basic_model2} is not identifiable in this general form given above because we may simply add some constant $c$ to $\alpha$ while at the same time subtracting $c$ from $f$. Then, the two models are equivalent. In other words, $\alpha$ and $f$ are only identifiable up to vertical shifts. That is why an overall constant $\alpha_0$ is typically introduced in terms of
\begin{equation}\label{eq:basic_alpha_seperated}
 u_j(t) = \alpha_0 + \tilde{\alpha}(t) + f(z_j(t)) + E_j(t),   
\end{equation}
and $\tilde{\alpha}$ and $f$ are centered in some sense. The constraint used in the R package \texttt{mgcv} \citep{Wood_2017}, which we use here for model training, is $\sum_{i,j} \tilde{\alpha}(t_{ji}) = 0 \; \mbox{ and } \sum_{i,j} f(z_j(t_{ji})) = 0$, respectively. In what follows, we will typically report the functional intercept $\alpha(t) = \alpha_0 + \tilde{\alpha}(t)$ and the centered $f$. Furthermore, it is worth noting that measurement points $t_{ji}$ neither need to be the same for each day nor on a regular grid. As a consequence, missing values in the output profiles or the covariate curves are allowed. 
%%%%%%%%%%%%%%%%%%%%%%%%%%%%%%%%%%%%%%%%%%%%%%%%%%%%%%%%%%%%%%%%%%%%
\subsubsection{Modified and extended models}\label{subsubsec:extend} 
%%%%%%%%%%%%%%%%%%%%%%%%%%%%%%%%%%%%%%%%%%%%%%%%%%%%%%%%%%%%%%%%%%%%
If model \eqref{eq:basic_model} is simplified, parameterization \eqref{eq:basic_model2} simplifies, as well. For instance, if an output-only approach based entirely on FPCA is used, the part in~\eqref{eq:basic_model2} referring to the regression function $f$ vanishes. If the latter is assumed to be linear, only the corresponding $\beta$ has to be estimated. If $\beta$ is allowed to change over the course of the day in terms of $\beta(t)$, this function can be expanded in basis functions as it was done with $f$ in \eqref{eq:f_basis} and estimated analogously. If, on the other hand, multiple covariates are given, and model \eqref{eq:basic_model} is extended to $\sum_{k=1}^p f_k(z_{jk}(t))$ instead of a single $f(z_j(t))$, we have to add basis representations 
$\sum_{l_k=1}^{L^{(f_k)}} \gamma^{(f_k)}_{l_k} b^{(f_k)}_{l_k}(z_{jk}(t))$ for each component $f_k$, $k=1,\ldots,p$, in \eqref{eq:basic_model2}. If we want to account for both daily and yearly patterns in a purely additive way, that is, $\alpha(t,d_j) = \alpha_0 + \tilde{\alpha}_1(t) + \tilde{\alpha}_2(d_j)$, the procedure is completely analogous. However, if $\alpha(t,d_j)$ is supposed to be a surface, or if a two-way interaction, say $f(z_{j1}(t),z_{j2}(t))$ in~\eqref{eq:basic_model}, is to be fitted, some minor modifications are necessary. The unknown functions ($\alpha$ and $f$, respectively) now have two arguments, meaning that the basis needs to be chosen appropriately, e.g., as a so-called tensor product basis. For instance, if considering $f$, equation~\eqref{eq:f_basis} turns into
\begin{equation}\label{eq:f_basis2}
    f(z_1,z_2) = \sum_{l_1=1}^{L_1}\sum_{l_2=1}^{L_2} \gamma_{l_1,l_2} b_{l_1}(z_1) b_{l_2}(z_2),
\end{equation}
and smoothing is typically done in both the $z_1$- and the $z_2$-direction. Having said that, the model complexity increases in terms of the number of basis coefficients $\gamma_{l_1,l_2}$ that need to be fitted. As pointed out earlier, higher-order interactions can also be estimated, but the number of known coefficients increases even further. Bivariate (or even higher dimensional) smoothers allowing for terms such as \eqref{eq:f_basis2} are available in \texttt{mgcv}, and corresponding estimates will be shown in the real-world data evaluations in Section~\ref{sec:Results}. The approach for bivariate $\alpha(t,d_j)$ or $f(z_j(t),t)$ cases, where the potentially nonlinear covariate effect is allowed to change over the course of the day, is analogous. In summary, in any of the cases considered here, the final model is linear in the basis coefficients and random effects, and all those quantities can be estimated/predicted through penalized least squares in a mixed model framework.
%%%%%%%%%%%%%%%%%%%%%%%%%%%%%%%%%%%%%%%%%%%%%%%%%%%%%%%%%%%%%%%%%%
\subsubsection{Estimation of eigenfunctions}\label{subsubsec:fpca}
%%%%%%%%%%%%%%%%%%%%%%%%%%%%%%%%%%%%%%%%%%%%%%%%%%%%%%%%%%%%%%%%%%
In Section \ref{subsubsec:basic} and \ref{subsubsec:extend}, 
we assumed that the eigenfunctions $\phi_r$, $r=1,\ldots,m$, are known. In practice, those functions need to be estimated from the training data in some way. To do so, we first fit model~\eqref{eq:basic_model}, or a modification from above, with a working independence assumption concerning $E_j(t)$ \citep{Scheipl.etal_2015}. Then, we use the resulting estimates of the error process for FPCA, plug in the estimated eigenfunctions $\hat{\phi}_r$, and fit the final model as discussed above. Such a two-step approach has already worked well in the past \citep{Gertheiss.etal_2017}. For FPCA, we use an approach based on \cite{Yao.etal_2005} that accommodates functions with additional white noise errors and thus works relatively generally \citep{Gertheiss.etal_2024}. The idea is to incorporate smoothing into estimating the covariance in \eqref{eq:mercer}. The estimation is based on the cross-products of observed points within error curves, $E_j(t_{ji})E_j(t_{ji'})$, which are rough estimates of $\Cov(E(t_{ji}), E(t_{ji'}))$. All cross-products are pooled, and a smoothing method of choice is used for bivariate smoothing. \cite{Yao.etal_2005} used local polynomial smoothing, while the applied \texttt{refund} R package \citep{Goldsmith.etal_2022} employs penalized splines. If an additional error is assumed (as we do here), the diagonal cross-products approximate the variance of the structural component plus the error variance. The diagonal is thus left out for smoothing. Once the smooth covariance is available, the orthogonal decomposition \eqref{eq:mercer} is done numerically on a fine grid using the usual matrix eigendecomposition.  

\begin{figure}[!htb]
\centering
\includegraphics[scale=.65]{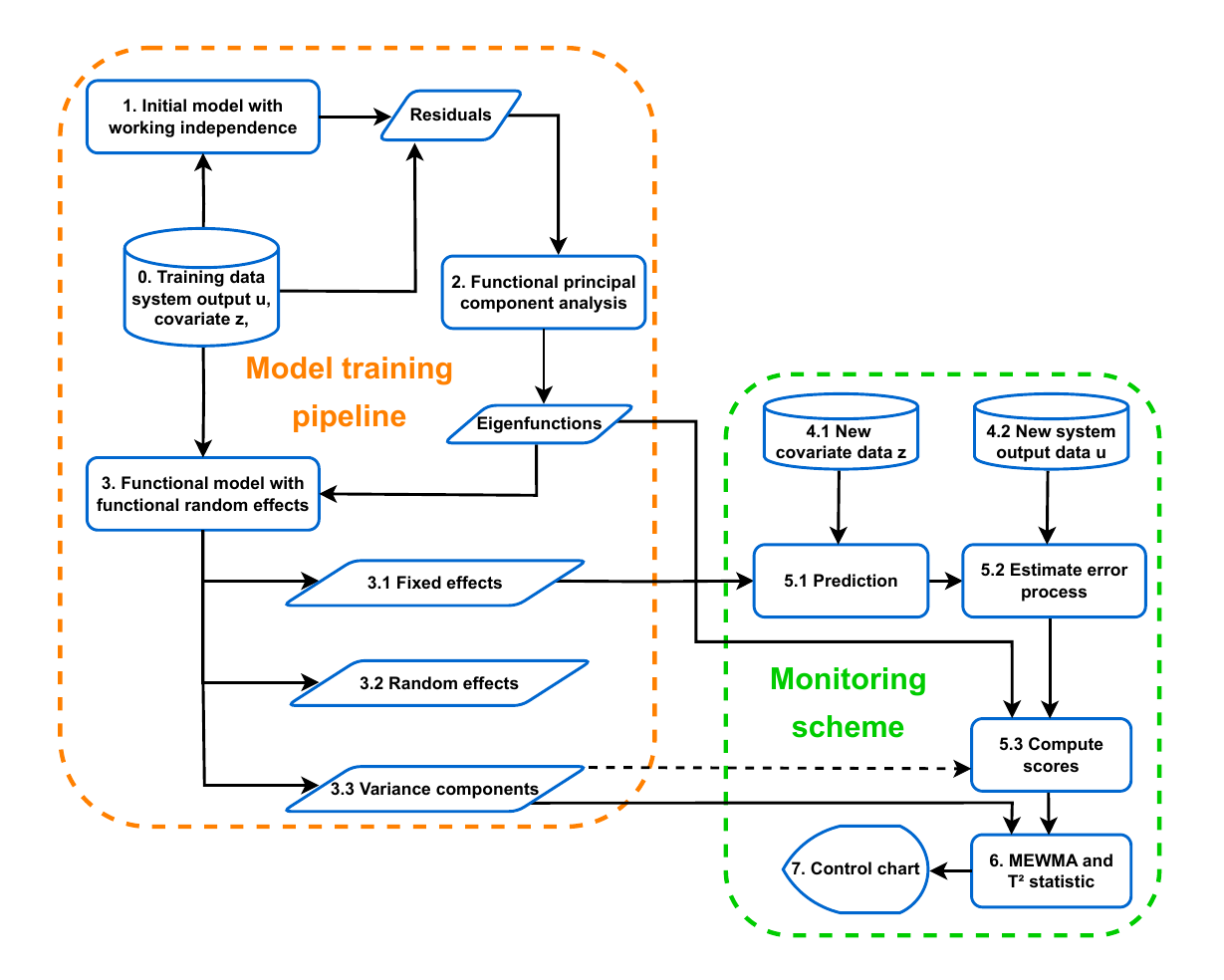}
\caption{Visual summary of the CAFDA-SHM framework.}
\label{fig:flowchart_procedure}
\end{figure}

The entire model training workflow is summarized in the orange part of Figure~\ref{fig:flowchart_procedure}. We first use the Phase-I data (step 0) to fit the initial model with working independence in step 1. On the resulting residuals, the ``observed'' error process, we carry out FPCA to obtain estimates of the eigenfunctions (step 2). For choosing the number of components, we use the usual threshold of 95\% or 99\% of the variance explained. Then, the eigenfunctions are used to train the final model in step 3. From this model, we obtain estimates of the fixed and random effects, the variance components, and the residuals. Particularly, the fixed effects, the eigenfunctions, and the variance components will be needed to estimate Phase-II component scores and implement a monitoring scheme as described below.
%%%%%%%%%%%%%%%%%%%%%%%%%%%%%%%%%%%%%%%%%%%%%%%%%%%%%%%%%%%%%%%%%%%%%
\subsubsection{Estimation of Phase-II scores}\label{subsubsec:scores}
%%%%%%%%%%%%%%%%%%%%%%%%%%%%%%%%%%%%%%%%%%%%%%%%%%%%%%%%%%%%%%%%%%%%%
Once the mixed model for function-on-function regression has been trained, it can be used to monitor future system outputs if the covariates in the trained model are available as well. The corresponding data is denoted as ``new covariate data $z$'' and ``new system output data $u$'' in Figure~\ref{fig:flowchart_procedure}. An essential input for the monitoring scheme described in the next subsection is the principal component scores $\xi_{1,g},\ldots,\xi_{m,g}$ for a new day $g$ in Phase-II. To estimate those scores from the new data, we first use the fixed effects from the Phase-I model, e.g., $\hat{\alpha}(t_{gi})$ and $\hat{f}(z_g(t_{gi}))$ in case of model~\eqref{eq:basic_model}, to obtain a prediction for the system outputs $u_g(t_{gi})$. Here, $t_{gi}$, $i=1,\ldots,N_g$, denote the time instances of the covariate $z_g(t_{gi})$ available at day $g$. Then, those predictions are subtracted from the actually observed $u_g(t_{gi})$ to obtain estimated measurements $\hat{E}_g(t_{gi})$ of the error process. If those time points are sufficiently dense over the course of day $g$, which is typically the case if day $g$ is (nearly) over, the scores can be estimated through numerical integration such that
\begin{equation}\label{eq:estSc1}
    \hat{\xi}_{r,g}= \int_\mathcal{T} \hat{E}_g(t) \hat{\phi}_r(t)dt.
\end{equation}
However, sometimes there is a substantial amount of missing values in the $u$ or $z$ data, e.g., due to technical problems, or --- more importantly --- the score estimates have to be available before the new day is over. In the latter case, we may say that all the data points after some time point are ``missing''. In both situations, the interpretation as a mixed model is helpful, as this also allows for predicting the random effects for a reduced set of measurement points $t_{gi}$.  

For that purpose, let $\boldsymbol{\phi}_{gr} = (\phi_r(t_{g1}),\ldots,\phi_r(t_{g N_g}))^\top$ be the $r$th eigenfunction evaluated at time points $t_{gi}$, $i=1,\ldots,N_g$, $r=1,\ldots,m$, and $\Sigma_{\mathbf{E}_g}$ the covariance matrix of the error vector $\mathbf{E}_g = (E_g(t_{g1}),\ldots,E_g(t_{g N_g}))^\top$. Then, assuming a Gaussian distribution, the conditional expectation of the score $\xi_{r,g}$ given $\mathbf{E}_g$ is  
$\mathbb{E}(\xi_{r,g}|\mathbf{E}_g) = \nu_r \boldsymbol{\phi}_{gr}^\top \Sigma_{\mathbf{E}_g}^{-1}\mathbf{E}_g, \;\; r=1,\ldots,m$ \citep{Yao.etal_2005}. Due to \eqref{eq:Ej1} and \eqref{eq:wj}, the matrix $\Sigma_{\mathbf{E}_g}$ can be estimated as $\hat{\Sigma}_{\mathbf{E}_g} = \hat{\boldsymbol{\Phi}}_g \mbox{diag}(\hat{\nu}_1,\ldots,\hat{\nu}_m) \hat{\boldsymbol{\Phi}}_g^\top + \hat{\sigma}^2\mathbf{I}_{N_g}$, with $\hat{\boldsymbol{\Phi}}_g = (\hat{\boldsymbol{\phi}}_{g1}|\ldots|\hat{\boldsymbol{\phi}}_{gm})$. After plugging in the estimates $\hat{\sigma}^2$, $\hat{\nu}_r$, $\hat{\phi}_r$, $r=1,\ldots,m$, from the model training phase and $\hat{\mathbf{E}}_g = (\hat{E}_g(t_{g1}),\ldots,\hat{E}_g(t_{g N_g}))^\top$ from above, we obtain the estimated scores
\begin{equation}\label{eq:estSc2}
    \hat{\xi}_{r,g} = \hat{\nu}_r \hat{\boldsymbol{\phi}}_{gr}^\top \hat{\Sigma}_{\mathbf{E}_g}^{-1}\hat{\mathbf{E}}_g, \;\; r=1,\ldots,m.
\end{equation}
These scores can then be used as input to a control chart. The workflow of estimating Phase-II scores is also summarized in Figure~\ref{fig:flowchart_procedure} (step 4.1~--~5.3). The dashed arrow towards the scores (5.3 in Figure~\ref{fig:flowchart_procedure}) indicates that variance components $\hat{\nu}_1,\ldots,\hat{\nu}_m$ and $\hat{\sigma}^2$ are needed if using \eqref{eq:estSc2} but not for \eqref{eq:estSc1}.
%%%%%%%%%%%%%%%%%%%%%%%%%%%%%%%%%%%%%%%%%%%%%%%%%%%%%%%%%%%%%%%%%%%%%
\subsection{Control charts}\label{sec:ControlCharts}
%%%%%%%%%%%%%%%%%%%%%%%%%%%%%%%%%%%%%%%%%%%%%%%%%%%%%%%%%%%%%%%%%%%%%
After accounting for environmental influences through appropriate modeling, the system outputs can be monitored by applying a control chart. A multivariate Hotelling control chart is often used to quickly detect a shift in the structural condition \citep{Deraemaeker.etal_2008, Magalhaes.etal_2012, Comanducci.etal_2016}. Here, we employ a memory-type control chart, the Multivariate Exponentially Weighted Moving Average (MEWMA) introduced by \cite{Lowry.etal_1992} to the Phase-II scores $\xi_{1,g},\ldots,\xi_{m,g}$, $g=1,2,\ldots$, from Section~\ref{subsubsec:scores}. The MEWMA contains the Hotelling chart as a special case. The MEWMA chart assumes serial independent normally distributed vectors $\boldsymbol{\xi}_1, \boldsymbol{\xi}_2, \dots$ of dimension $m$ with $\boldsymbol{\xi}_g\sim\mathcal{N}(\boldsymbol{\mu}, \boldsymbol{\Lambda})$
where the scores $\xi_{r,g}$ from \eqref{eq:estSc1} or \eqref{eq:estSc2} are used in $\boldsymbol{\xi}_g$ such that $\boldsymbol{\xi}_g = (\xi_{1,g},\ldots,\xi_{m,g})^\top$. We follow \cite{Knoth_2017} and define a mean vector $\boldsymbol{\mu}$ that follows a change point model $\boldsymbol{\mu}=\boldsymbol{\mu}_0$ for $g<\tau$ and $\boldsymbol{\mu}=\boldsymbol{\mu}_1$ for $g\geq\tau$ for an unknown time point $\tau$ and by definition  $\boldsymbol{\mu}_0=\boldsymbol{0}$, see Section~\ref{subsec:model}. For in-control data, the scores $\xi_{1,g},\ldots,\xi_{m,g}$ are assumed to be uncorrelated with variances $\nu_1,\ldots,\nu_m$, see also Section~\ref{subsec:model}. Hence, the covariance matrix $\boldsymbol{\Lambda}$ is diagonal with $\boldsymbol{\Lambda}=\text{diag}(\nu_1, \dots, \nu_m)$. Then, we apply the following smoothing procedure to compute the MEWMA statistic (step 6 in Figure \ref{fig:flowchart_procedure})
\begin{equation}\label{eq:MEWMA}
 \boldsymbol{\omega}_g=(1-\lambda)\boldsymbol{\omega}_{g-1} + \lambda\boldsymbol{\xi}_g, \quad  \boldsymbol{\omega}_0=\boldsymbol{0}\
\end{equation}
with $g=1, 2, \dots$ and smoothing constant $0 < \lambda \leq 1$. The smoothing parameter $\lambda$ controls the sensitivity of the shift to be detected. Smaller values of $\lambda$ such as $\lambda \in \{0.1, 0.2, 0.3\}$, are usually selected to detect smaller shifts \citep{Hunter_1986}, while $\lambda=1$ results in the Hotelling chart. In this study, we use $\lambda=0.3$. The control statistic is the Mahalanobis distance
\begin{equation}\label{eq:MHD}
T^2_g=(\boldsymbol{\omega}_g-\boldsymbol{\mu}_0)^\top\boldsymbol{\Lambda}^{-1}_{\boldsymbol{\omega}}   (\boldsymbol{\omega}_g-\boldsymbol{\mu}_0),
\end{equation}
with asymptotic covariance matrix of $\boldsymbol{\omega}_g$,   $\boldsymbol{\Lambda}_{\boldsymbol{\omega}}=\lim_{g\rightarrow\infty} \text{Cov}(\boldsymbol{\omega}_g) = \Big\{\frac{\lambda}{2-\lambda}\Big\}\boldsymbol{\Lambda}$.

The MEWMA chart issues an alarm if $T_i^2>h_4$, i.e., the control statistic is above the threshold value $h_4$. The stopping time $N=\inf{\{g \geq 1:T^2_g> h_4\}}$, also known as average run length (ARL), is often used to measure the control chart's performance. It is defined as the average number of observations until the chart signals an alarm. If the process is in-control, the ARL (ARL$_0$) should be high to avoid false alarms. If there is a change in the underlying process, the ARL (ARL$_1$) should be low to detect changes quickly. To determine the threshold value, the ARL must be calculated when the process is in-control, usually applying a grid search or a secant rule. This ARL$_0$ can be calculated as described in \cite{Knoth_2017} and is implemented in R-package \texttt{spc} \citep{Knoth_2022}. An evaluation of our CAFDA framework on artificial data is demonstrated in the next section.

%%%%%%%%%%%%%%%%%%%%%%%%%%%%%%%%%%%%%%%%%%%%%%%%%%%%%%%%%%%%%%%
\section{Illustration on Artificial Data}\label{sec:Simulation}
%%%%%%%%%%%%%%%%%%%%%%%%%%%%%%%%%%%%%%%%%%%%%%%%%%%%%%%%%%%%%%%
The aim of this section is to validate the methodology proposed in Section \ref{sec:FDA} on artificial data in a Monte Carlo simulation study. To this end, we first describe the data generating process (DGP) of the functional data in the following Section~\ref{subsec:data_generation}. We then apply the modeling part of our framework to the artificially generated data in Section~\ref{subsec:model_training_results} to provide insight into the modeling performance. In the last Section~\ref{subsec:monitoring_results}, we evaluate the proposed monitoring scheme (the ``green part'' in Figure \ref{fig:flowchart_procedure}) in the CAFDA-SHM framework.
%%%%%%%%%%%%%%%%%%%%%%%%%%%%%%%%%%%%%%%%%%%%%%%%%%%%%%%%%%
\subsection{Data generation}\label{subsec:data_generation}
\begin{figure}[!htb]
\centering
\includegraphics[scale=.5]{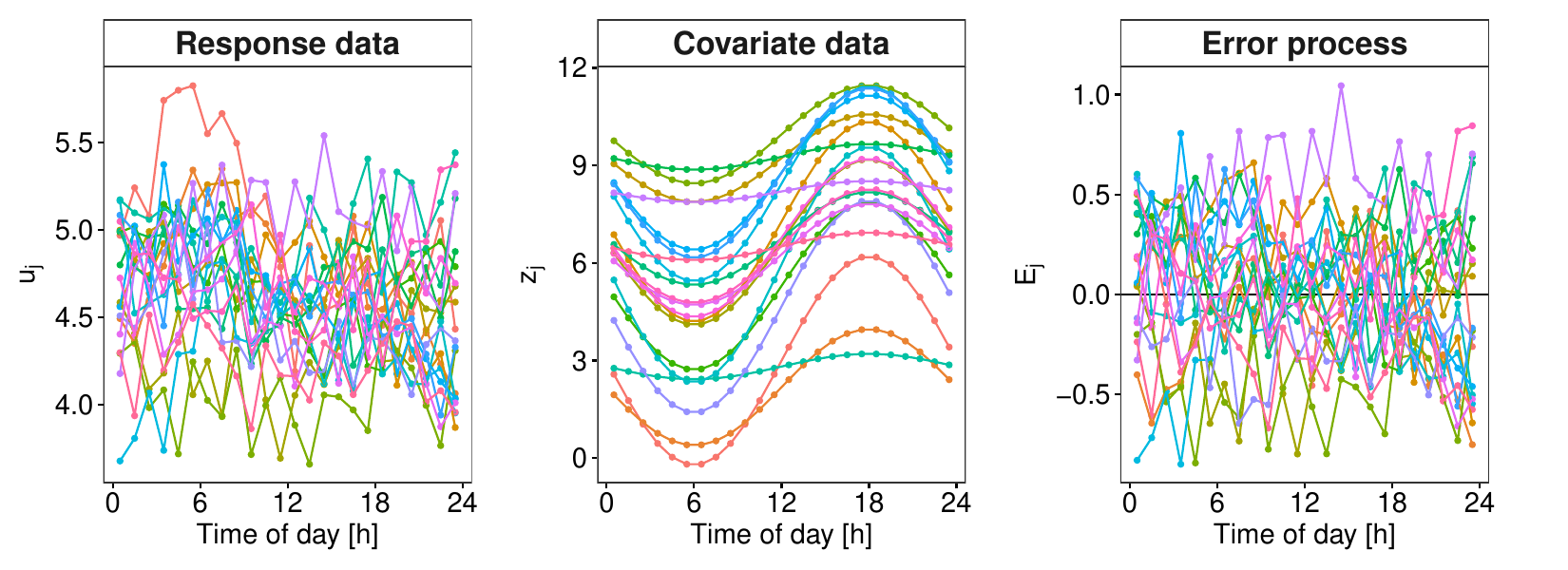}
\caption{Sample of 20 generated output profiles (left), covariate profiles (middle), and error profiles (right).}
\label{fig:simulation_sample_profiles}
\end{figure}

%%%%%%%%%%%%%%%%%%%%%%%%%%%%%%%%%%%%%%%%%%%%%%%%%%%%%%%%%%
We consider the basic model~\eqref{eq:basic_model} with one system output and one covariate. To generate the data, we consider each component of the model separately. The eigenfunctions $\phi_r$, $r=1,2,3$, are obtained using orthonormal Legendre Polynomials \citep{Abramowitz.Setgun_1964} up to an order of two.
Following \eqref{eq:Ej1}, we split $E_j$ into two parts. First, the structural component $w_j(t)$, $t \in \mathcal{T}$, $ \mathcal{T} = (0, 24)$, is generated by $w_j(t)=\xi_{1j}\phi_1(t) + \xi_{2j}\phi_2(t) + \xi_{3j}\phi_3(t),$
where the individual scores are obtained through $\xi_{r,j}\overset{\text{iid}}{\sim}\mathcal{N}(0, \nu_r)$ $r=1,2,3$, $j=1,\dots, J$. The eigenvalues $\nu_r=\exp(-\frac{r+1}{2})$ are decreasing exponentially towards zero \citep{Happ-Kurz_2020}. The second component of $E_j$, the white noise at the measurement points $t_i$, $i=1,\ldots,24$, is generated by $\epsilon_j(t_i) \overset{\text{iid}}{\sim} \mathcal{N}(0, 0.2)$. A functional intercept $\alpha$ is also created, separated into $\alpha_0$ and $\tilde{\alpha}(t)$. While the functional component is $\tilde{\alpha}(t)=\sin\big(\frac{\pi t}{48}\big)+\cos\big(\frac{\pi t}{6}\big)$, $t \in \mathcal{T}$, and centered by subtracting its mean value, the overall intercept is set to $\alpha_0=5$. To form a functional covariate that can represent different shapes of cyclic daily profiles analogous to temperature profiles at different times of the year, a sine function in $z_j(t) = \zeta_{0j}+\zeta_{1j}\sin\big(\frac{\pi t}{12}+0.3\big)$, $t \in \mathcal{T}$, with $\zeta_{0j}, \zeta_{1j}\overset{\text{iid}}{\sim} U(a,b)$ is used, where $U(a,b)$ denotes the uniform distribution over the interval $(a,b)$. Thus, the covariate profile level is set by $\zeta_{0j}$ with $a=2$ and $b=12$, and its shape by $\zeta_{1j}$, here with $a=0$ and $b=4$. Subsequently, the covariate data $z_j(t)$ is transformed by the regression function $f(z)=\exp{\big(-\frac{11}{5}z\big)}-0.5$ and the output $u_j$ for day $j$ at the measurement points $t_i$ is generated through $u_j(t_i) = \alpha_0 + \tilde{\alpha}(t_i) + f(z_j(t_i)) + w_j(t_i) + \epsilon_j(t_i) $, see Equations \eqref{eq:Ej1} and \eqref{eq:basic_alpha_seperated}. Figure \ref{fig:simulation_sample_profiles} shows a sample of 20 generated profiles of the output $u_j$, the covariate $z_j$, and the error process $E_j$ similar to the real data in Figure \ref{fig:KW51_bilinear}. For our evaluations below, the sample size was $J=300$.

%%%%%%%%%%%%%%%%%%%%%%%%%%%%%%%%%%%%%%%%%%%%%%%%%%%%%%%%%%%%%%%%%%%%%%%%%%%%%%%%%%%
\subsection{Model training simulation results}\label{subsec:model_training_results}
\begin{figure}[!htb]
\centering
\includegraphics[scale=.62]{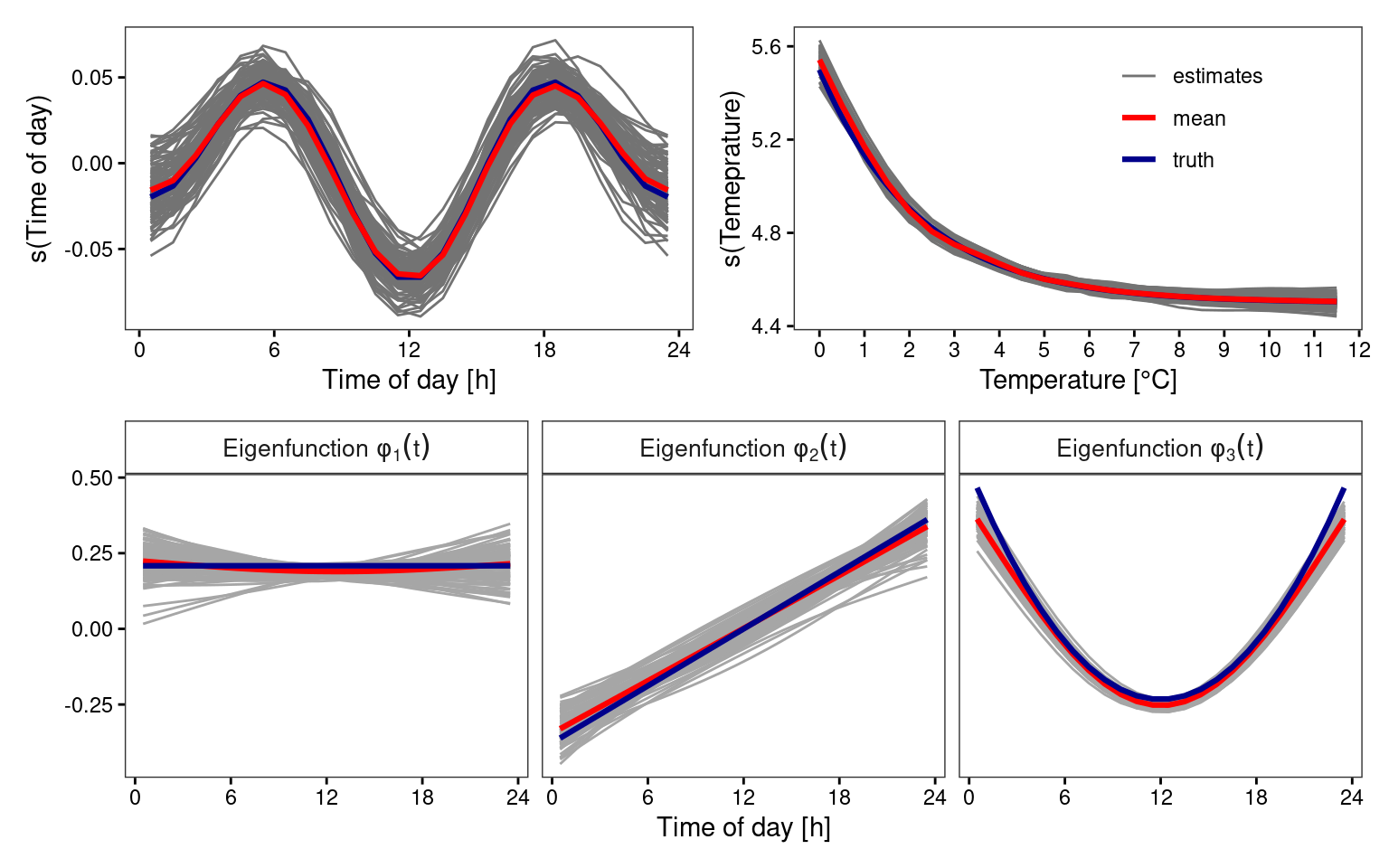}
\caption{The estimates of the functional intercept (top-left) and the nonlinear covariate effect (top-right), as well as the estimated eigenfunctions of the structural component of the error process (bottom row). Shown in grey are the results for each of the 100 simulation runs, while the blue curves give the true functions, and the red curves the mean across the 100 runs.}
\label{fig:simulation_results_modeling}
\end{figure}

%%%%%%%%%%%%%%%%%%%%%%%%%%%%%%%%%%%%%%%%%%%%%%%%%%%%%%%%%%%%%%%%%%%%%%%%%%%%%%%%%%%
Next, we perform the model training pipeline of the CAFDA-SHM framework (steps 0 to 3, the ``orange part'' in Figure \ref{fig:flowchart_procedure}). We repeated the procedure 100 times to gain insights into the model performance and possible parameter estimation variability. Figure \ref{fig:simulation_results_modeling} illustrates the results comparing the true functions $\tilde{\alpha}(t)$, $\alpha_0 + f(z)$, and $\phi_r(t)$, $r=1,\ldots,3$, drawn in blue, to the estimates $\hat{\tilde{\alpha}}(t)$, $\hat\alpha_0 + \hat{f}(z)$, $\hat{\phi}_r(t)$ from the 100 simulation runs (grey), where each run consisted of a training set of $J=300$ $u$- and $z$-profiles. The respective mean functions of the estimates are shown in red. The top-left figure shows the results for the centered functional intercept $\tilde{\alpha}(t)$ and the top-right for the covariate effect plus the overall intercept $\alpha_0 + f(z)$. The three figures in the bottom row show the results for eigenfunctions $\phi_r(t)$, $r=1,2,3$. The red and blue curves are close, meaning that the trained models and their estimated functional components can approximate the true underlying functions well on average. In other words, the estimates are (nearly) unbiased. Also, the variation is relatively small, see the grey curves.
%%%%%%%%%%%%%%%%%%%%%%%%%%%%%%%%%%%%%%%%%%%%%%%%%%%%%%%%%%%%%%%%%%%%%%%%%%%
\subsection{Monitoring simulation results}\label{subsec:monitoring_results}
%%%%%%%%%%%%%%%%%%%%%%%%%%%%%%%%%%%%%%%%%%%%%%%%%%%%%%%%%%%%%%%%%%%%%%%%%%%
In this section, the monitoring scheme (green part in Figure \ref{fig:flowchart_procedure}) is evaluated concerning the out-of-control ARL performance in a Monte Carlo simulation with 10,000 replications. For this purpose, the DGP presented in \ref{subsec:data_generation} is used, as well as all 100 models with $J=300$ and its estimated fixed effects and variance components in \ref{subsec:model_training_results}.
The control limit $h_4$ is determined based on prespecified ARL$_0=100$, and the ARL$_1$ is computed by the following procedure for Phase-II days $g=1,2,\ldots$:
\begin{enumerate}
    \item Generate new covariate data $z_g(t)$ according to Section~\ref{subsec:data_generation} (which produces the data in step 4.1 of Figure \ref{fig:flowchart_procedure}).
    \item Add a shift $\delta$ in one of the three components of $\boldsymbol{\mu}_0$ to create $\boldsymbol{\mu}_1$ and draw scores from the corresponding normal distribution to simulate the functional random effect $w_g(t)$.
    \item Generate the output data $u_g(t_i)$ at measurement points $t_1,\ldots,t_{24}$ by adding the functional intercept, the fixed effect of temperature $f(z_g(t_i))$, the functional random effect $w_g(t_i)$, and white noise error $\epsilon_g(t_i)$, analogously to Section~\ref{subsec:data_generation}. This produces the data in step 4.2 of Figure \ref{fig:flowchart_procedure}.
    \item Predict the system output using the model from Phase-I (compare Section~\ref{subsec:model_training_results}) and compute the resulting $\hat{E}_g(t_i)$ by subtracting the predicted from the observed system output. 
    \item Calculate the scores $\hat{\xi}_{r,g}$ according to Eq.~\eqref{eq:estSc2}.
    \item Compute MEWMA and $T^2$ Statistic according to Eq.~\eqref{eq:MEWMA} and Eq.~\eqref{eq:MHD}.
    \item If $T^2>h_4$, note the run-length $N$, else repeat steps 1.~-~6.~for the next profile (day).
\end{enumerate}

\begin{figure}[!htb]
\centering
\includegraphics[scale=.57]{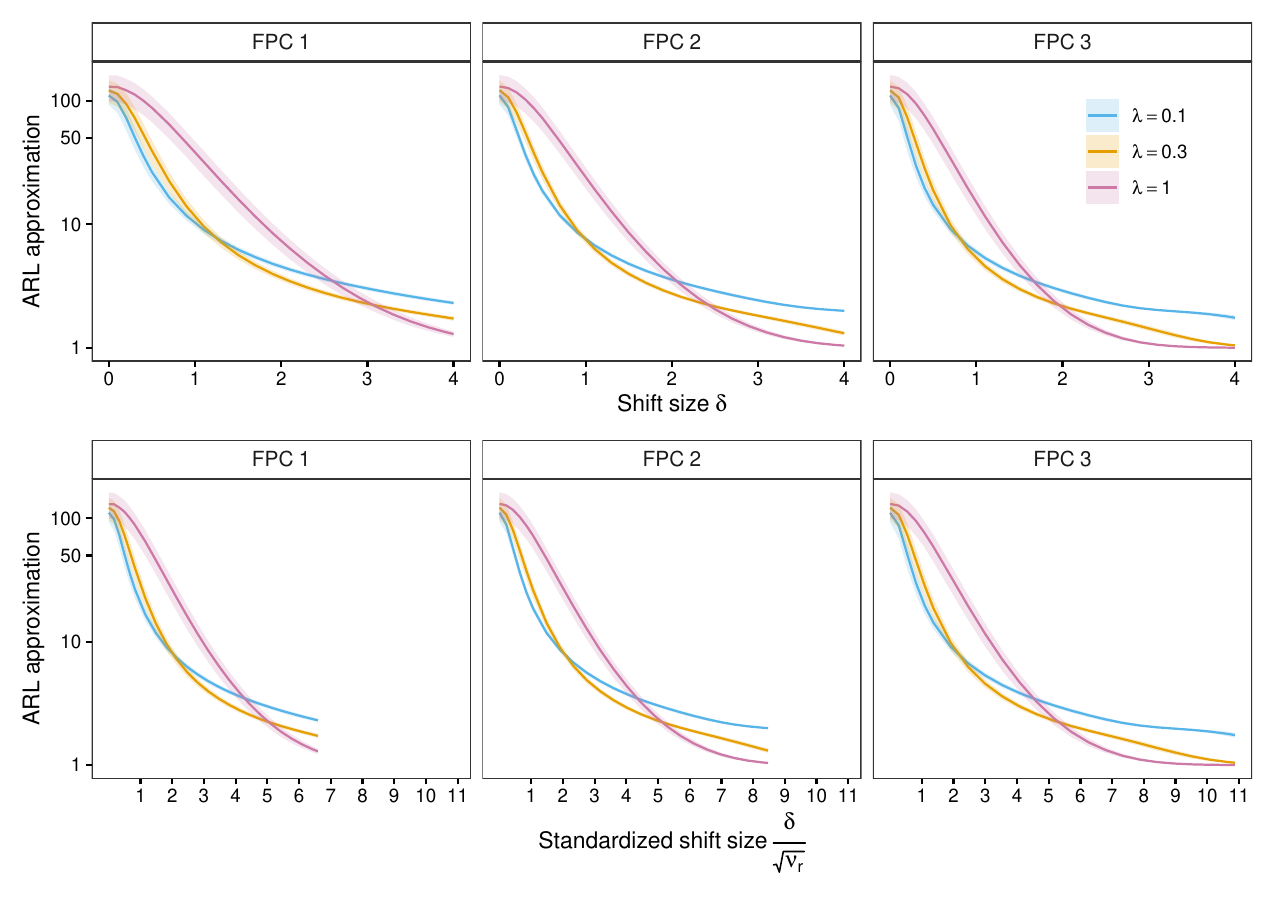}    
\caption{ARL profiles of MEWMA control charts on a logarithmic scale for a shift in the scores' mean with different smoothing parameter $\lambda$ displayed for shift size $\delta$ (top row) and standardized shift size $\frac{\delta}{\sqrt{\nu_r}}$ (bottom row). The model uncertainty is highlighted by the computed average ARL of all 100 models and a one standard deviation interval.}
\label{fig:simulation_results_monitoring}
\end{figure}

The upper panel of Figure \ref{fig:simulation_results_monitoring} shows ARL profiles on a logarithmic scale for shifts in the scores' mean on a fine grid for $\delta \in [0,4]$ with three different smoothing parameters $\lambda \in \{0.1, 0.3, 1\}$. For a fair comparison, all three charts are calibrated to the same $\text{ARL}_0=100$. However, note that the ARL of all three charts is slightly above 100 for the in-control scenario. This can be explained by the fact that the prediction (11) of the scores in the mixed model framework imposes a Ridge-type penalty and, hence, a form of shrinkage toward zero. Furthermore, it can be seen that the choice of the smoothing parameter $\lambda$ influences the ARL$_1$ profiles and, thus, the detection speed of the out-of-control scenario. As mentioned in Section \ref{sec:ControlCharts}, small values of $\lambda$ lead to quicker detection of small changes, while charts using, e.g., $\lambda=1$ (the pink curves in Figure~\ref{fig:simulation_results_monitoring}) have a shorter run length on average for substantial shifts. Comparing, for example, the crossing points of the ARL profiles for different shift sizes $\delta$ for the three components in the upper row, one could conclude that a shift in the third score would be detected more quickly than in the first component. However, this effect is blurred as the variances $\nu_r$ of the principal components decrease for larger $r$, see Section \ref{subsec:data_generation}. Hence, a version with standardized shift size $\frac{\delta}{\nu_r}$ is plotted in Figure \ref{fig:simulation_results_monitoring} (lower row) to give a clearer picture of the detection speed in regards to each component. It shows that the detection speed is comparable for all three components.

%%%%%%%%%%%%%%%%%%%%%%%%%%%%%%%%%%%%%%%%%%%%%%%%%%%%%%%%%%%%%%%%%%%%%%%%%%%%%%
\section{Applications to Structural Health Monitoring data}\label{sec:Results}
In what follows, we will apply our CAFDA-SHM framework to the data from the KW51 bridge in a first case study in Section \ref{sec:KW51}. We focus on the system's dynamic response, specifically the natural frequencies. Section \ref{sec:case_study_sachsengraben} provides a second case study on the strain measurements from a reinforced concrete motorway bridge, the Sachsengraben viaduct.
%%%%%%%%%%%%%%%%%%%%%%%%%%%%%%%%%%%%%%%%%%%%%%%%%%%%%%%%%%%%%%%%%%%%%%%%%%%%%%
\subsection{The KW 51 railway bridge}\label{sec:KW51}
 KW51 is a steel railway bridge of the bowstring type, with two curved, ballasted electrified tracks. It is 115 meters long, 12.4 meters wide, and located between Leuven and Brussels, Belgium, on the railway line L36N. The bridge was monitored from October 2nd, 2018, to January 15th, 2020, with a retrofitting period from May 15th to September 27th, 2019. Various quantities, such as the steel surface temperature or relative humidity, were measured on an hourly basis \citep{Maes.Lombaert_2020, Maes.Lombaert_2021, Maes.etal_2022}.  Due to the large amounts of data, only fragments of the raw acceleration and inclination data are available for re-use/download \citep{Maes.Lombaert_2020, Maes.Lombaert_2021} employed operational modal analysis (OMA) to determine the modal parameters with the reference-based covariance-driven stochastic subspace (SSI-cov/ref) algorithm on an hourly basis as well. The natural frequencies of 14 modes are available for the entire monitoring period and described in detail in \cite{Maes.Lombaert_2021}. Here, we will focus on the data for mode 6, with parts of it already shown in Figure~\ref{fig:KW51_bilinear}. In addition, we will consider two potentially confounding variables, temperature at the bridge deck level and relative humidity, both measured directly at the bridge. In a preprocessing step, some extreme outliers were removed from the data set. Those outliers correspond to some data points that resulted from abnormal bridge behavior on particularly cold days \citep{Maes.Lombaert_2021, Maes.etal_2022}. Eventually, 225 daily profiles were available for modeling before the retrofitting started. In Sections~\ref{subsec:basicmodelresults}--\ref{subsubsec:extended_model_results} below, we will discuss the CAFDA results for different model specifications, including the basic model with one covariate ``temperature'' only, an output-only version without any covariate information included, and two versions of an extended model with both temperature and relative humidity included as covariates. All our input-output models allow for nonlinear covariate effects, whereas earlier analyses of the KW51 data \citep{Anastasopoulos.etal_2021, Maes.etal_2022} focused on linear modeling. Finally, some results for monitoring Phase-II data will be shown in Section~\ref{subsubsec:kw51_monitoring}. To prevent the end of the in-control phase from coinciding with the start of the retrofitting, Phase I is limited to 200 observed profiles (between October 2, 2018, and April 19, 2019) for the model training. Out of those 200 profiles of natural frequencies, 129 profiles are observed for each of the 24 hours of the day, while the rest have between 4\% and 100\% of missing values. Concerning the temperature and humidity curves, 57 and 66 profiles have 4\%--100\% missing values, respectively. Overall, the percentage of missing values is high, with 63\%. 
%%%%%%%%%%%%%%%%%%%%%%%%%%%%%%%%%%%%%%%%%%%%%%%%%%%%%%%%
\subsubsection{Results for the basic model}\label{subsec:basicmodelresults}
%%%%%%%%%%%%%%%%%%%%%%%%%%%%%%%%%%%%%%%%%%%%%%%%%%%%%%%%
We start with the basic model and the parametrization in~\eqref{eq:basic_alpha_seperated}. It includes an overall intercept $\alpha_0$, a functional intercept $\tilde{\alpha}(t)$, a potentially nonlinear temperature effect $f(z)$, and the structural component $w_j(t)$ of the error process. After estimating the fixed effects of the initial model in step 1 of Figure \ref{fig:flowchart_procedure}, we apply the FPCA on $148$ residual profiles (those with sufficient data points of both the natural frequencies and temperature curves available) to extract the eigenfunctions of the functional random effects as described in Section~\ref{subsubsec:fpca}. The number of components to extract is chosen such that 99\% of the variance with respect to the smoothed covariance matrix is explained. In step 3 of our modeling framework, we refit the basic model, incorporating the eigenfunctions to account for the functional random effects.

\begin{figure}[!htb]
\centering
\includegraphics[scale=.46]{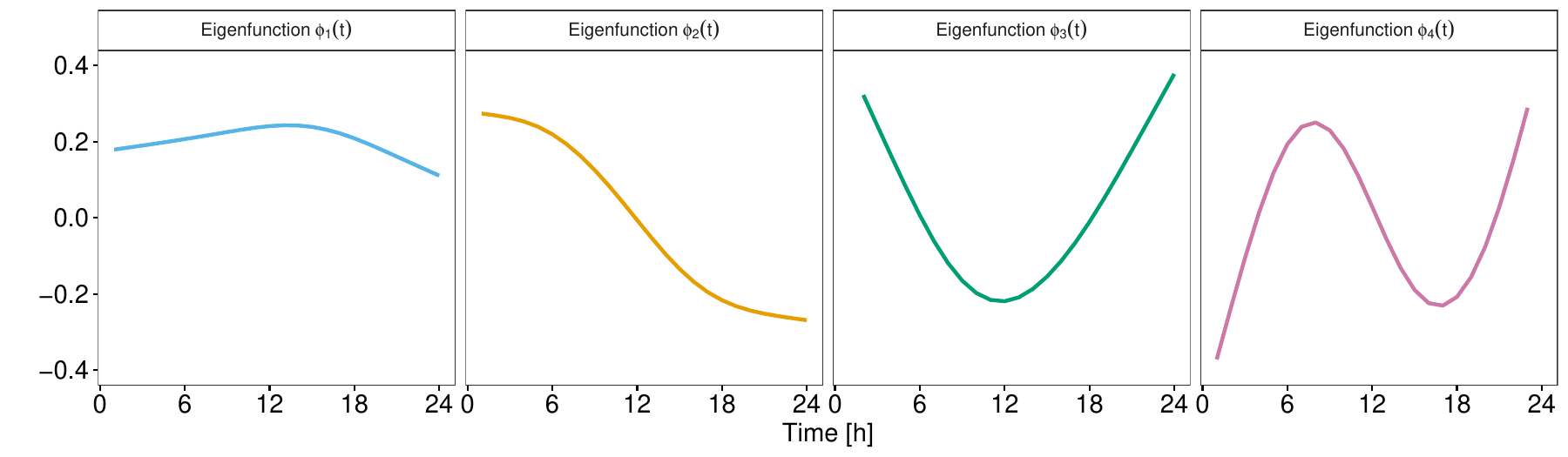}
\caption{Estimates of the eigenfunctions of the functional random effects in the basic model for the KW51 data.}
\label{fig:KW51_basic_model_eigenfunctions}
\end{figure}

Figure \ref{fig:KW51_basic_model_eigenfunctions} shows the estimates $\hat{\phi}_1(t),\ldots,\hat{\phi}_4(t)$ of the eigenfunctions. Note that the first two eigenfunctions already explain about 85.6\% of the variance of the structural part $w_j(t)$ of the error process. The first three eigenfunctions have straightforward and physically interpretable shapes. The first eigenfunction has (almost) the shape of a horizontal line, which means that the first principal component basically represents the overall extent of the daily error, indicating a slightly increased error variance in the early afternoon. The second component describes the difference in the errors between the morning and evening hours, whereas the third component corresponds to the contrast between night and day. The last component, component 4, is less clear, but this component only accounts for less than 5.6\% of the variance of $w_j(t)$. 

\begin{figure}[!htb]
\centering
\includegraphics[scale=.5]{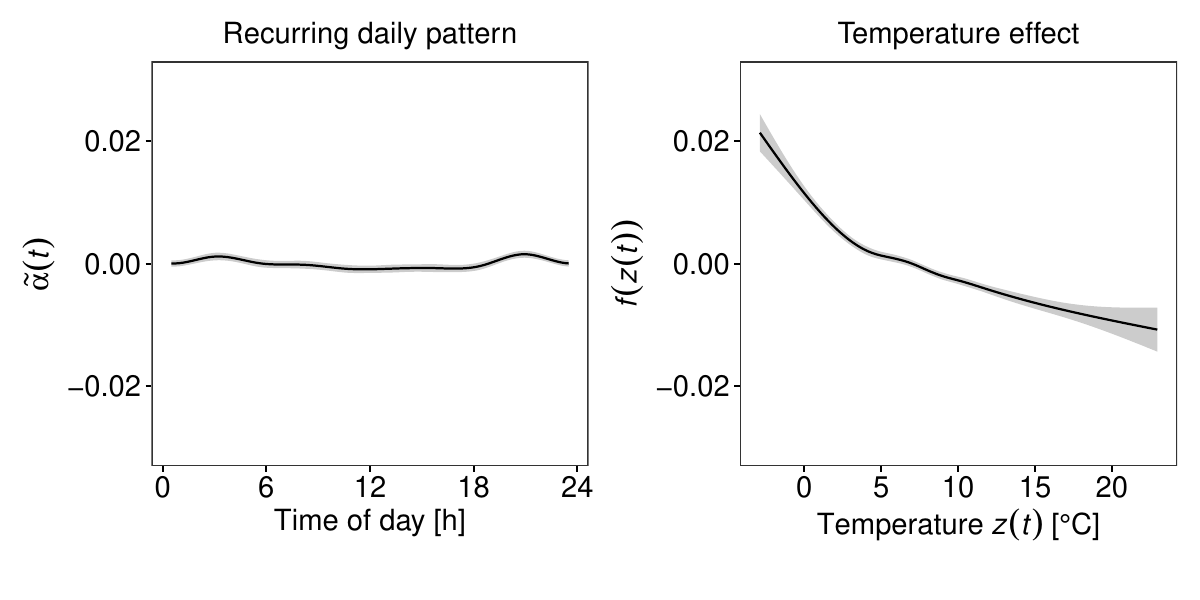}
\caption{Results of the basic functional modeling approach \eqref{eq:basic_model} for the functional intercept (left) and the nonlinear effect of temperature (right) on the natural frequency (mode 6) of the KW51 bridge.}
\label{fig:KW51_basic_model_estimated_effects}
\end{figure}

Figure \ref{fig:KW51_basic_model_estimated_effects} shows the centered functional intercept $\tilde{\alpha}$ and the nonlinear fixed effect $f(z)$ of temperature $z$ on the natural frequency of mode 6, which is also centered across the data observed here. The functional intercept can be interpreted as a recurring daily pattern. The grey-shaded areas represent the estimated effects' uncertainty in terms of pointwise 95\% confidence intervals. Comparing both effects, we find that the effect of the functional intercept is flat over the day, which means that a recurring daily pattern, if any, is very weak if the steel temperature is taken into account. In contrast, the temperature effect shows a pronounced nonlinear shape with a kink between 2°C and 3°C. This confirms statements in the literature that the influence of temperature on the dynamic response is stronger at lower temperatures \citep{Xia.etal_2012, Han.etal_2021}. Figure \ref{fig:KW51_basic_model_estimated_effects}~(right) also shows another benefit of the nonlinear approach compared to classical linear regression by \cite{Maes.etal_2022}, who simply omitted the data below 2°C. Our method can also utilize the data from the colder days and thus capture the nonlinear temperature effect well.
%%%%%%%%%%%%%%%%%%%%%%%%%%%%%%%%%
\subsubsection{Output-only model}
\begin{figure}[!htb]
\centering
\includegraphics[scale=.65]{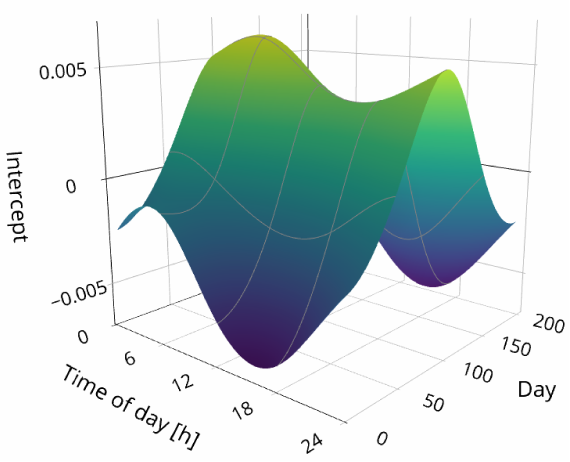}
\caption{The two-dimensional functional intercept in the output-only version (see Section \ref{subsec:model}) of the functional model for the natural frequency (mode 6) of the KW51 bridge.}
\label{fig:KW51_reduced_model_estimated_effect}
\end{figure}

%%%%%%%%%%%%%%%%%%%%%%%%%%%%%%%%%
As described in Section~\ref{subsec:model}, our model framework contains an output-only method as a special case in which only the system output of interest and its recorded 
timestamp are used, but no covariate data has to be recorded. Thus, we can use all available data for mode 6 in Phase I without being limited by missing covariate information, see Table \ref{tab:R2_of_models}. For modeling, we omit $f(z_j(t))$ but include an overall intercept $\alpha_0$ and a two-dimensional surface $\tilde{\alpha}(t, d_j)$ which considers seasonal patterns. Similar to the basic model, we perform steps 1--3 from Figure~\ref{fig:flowchart_procedure}. 
Figure \ref{fig:KW51_reduced_model_estimated_effect} visualizes $\tilde{\alpha}(t, d_j)$ of the final model. It is visible that the range of the functional intercept $\tilde{\alpha}(t, d_j)$ is much wider than the maximum effect sizes for the functional intercept in Figure \ref{fig:KW51_basic_model_estimated_effects}~(left). Apparently, the missing information on the temperature is now partly contained in the surface in Figure~\ref{fig:KW51_reduced_model_estimated_effect}. We can clearly see a daily and yearly pattern, where lower temperatures during the night and lower temperatures in the winter, respectively, lead to higher natural frequencies on average. More generally speaking, by specifying a two-dimensional functional intercept, we can already account for some of the environmental effects without the need to record corresponding covariate information. Consequently, in our opinion, such a functional intercept should always be included if training data is available over a sufficiently long period and an output-only approach is used to adjust for environmental influences. 

%%%%%%%%%%%%%%%%%%%%%%%%%%%%%%%%%%%%%%%%%%%%%%%%%%%%%%%%%%%%%%%%%%%%%%%
\subsubsection{Extended models}\label{subsubsec:extended_model_results}
\begin{figure}[!htb]
\centering
\includegraphics[scale=.45]{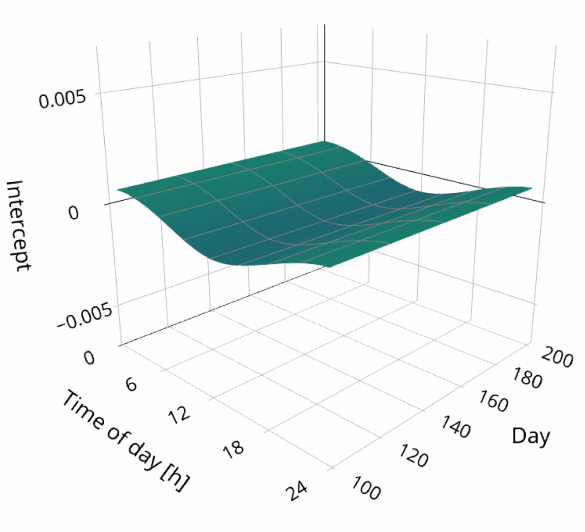}
\includegraphics[scale=.45]{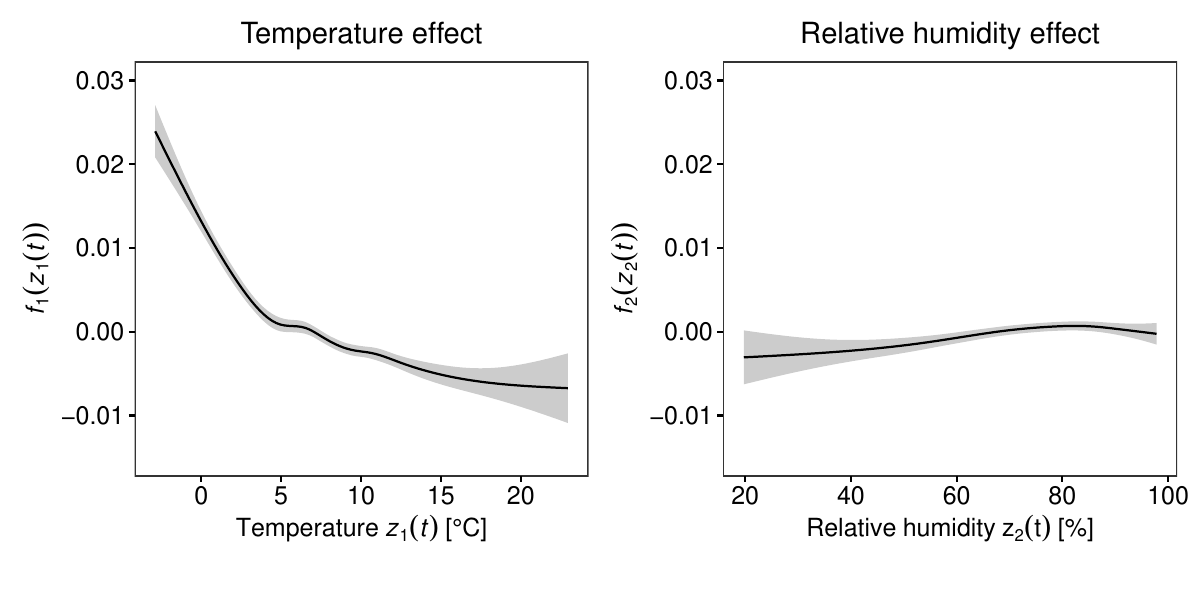}
\caption{Results of the extended functional modeling approach showing a two-dimensional functional intercept (left) and the additive, potentially nonlinear effects of temperature (middle) and relative humidity (right) on the natural frequency (mode 6) of the KW51 bridge.}
\label{fig:KW51_extended_model_additive_effects}
\end{figure}

%%%%%%%%%%%%%%%%%%%%%%%%%%%%%%%%%%%%%%%%%%%%%%%%%%%%%%%%%%%%%%%%%%%%%%%
Next, we extend the basic model by including relative humidity as a second covariate in an additive way. We now have $u_j(t) = \alpha_0 + \tilde{\alpha}(t, d_j) + f_1(z_1(t)) + f_2(z_2(t)) + E_j(t)$, $j=1, \ldots, J$, $t \in \mathcal{T}$, where $f_1$ and $f_2$ denote the potentially nonlinear effects of temperature and relative humidity, respectively. Due to the limited number of data sets where the two covariates are measured at the same time, the number of daily profiles that can be used for parameter estimation reduces to $J=100$ days. As before, we proceed with the model training steps 1--3 from Figure~\ref{fig:flowchart_procedure}. Figure~\ref{fig:KW51_extended_model_additive_effects} shows the results for the final model, the (centered) two-dimensional functional intercept (left), and the effects of temperature (middle) and relative humidity (right). 
The daily and yearly pattern captured by the intercept now only shows a minimal effect, which makes sense since the effects seen in Figure~\ref{fig:KW51_reduced_model_estimated_effect} were (presumably) mainly due to varying temperatures. If the latter is included explicitly in the model, those effects disappear from the functional intercept. The temperature effect $f_1$ shows a similar nonlinear behavior as seen with the basic model in Figure \ref{fig:KW51_basic_model_estimated_effects}. The effect $f_2$ for relative humidity is much smaller but still present, as can also be seen from the increased $R^2$ of $0.52$, compared to 0.43 for the basic model with temperature as the only covariate, see Table~\ref{tab:R2_of_models}. 

\begin{figure}[!htb]
\centering
\includegraphics[scale=.55]
{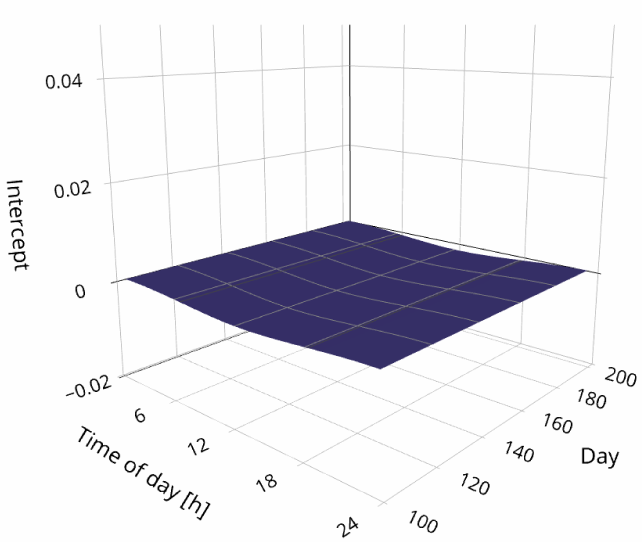}
\includegraphics[scale=.53]
{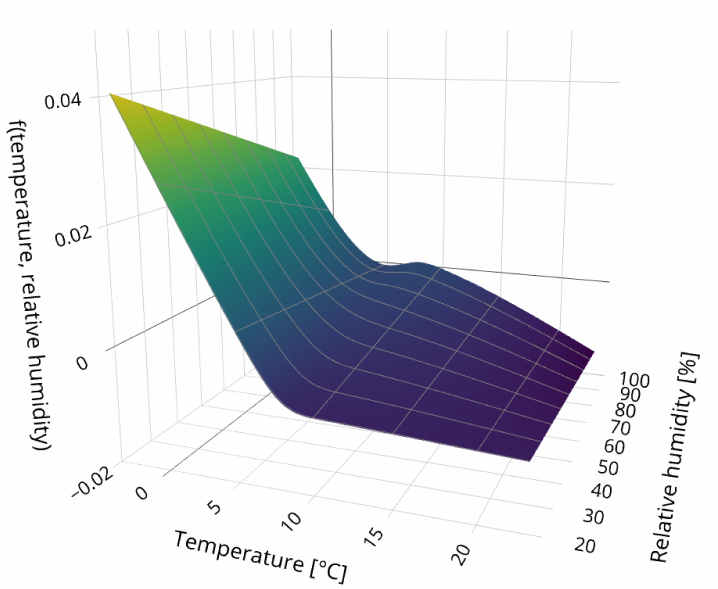}
\caption{Results of the extended functional modeling approach showing a two-dimensional functional intercept $\alpha(t, d{_j})$ (left) and a two-dimensional functional interaction of temperature and relative humidity $f_{12}(z_{j1}(t),z_{j2}(t))$ (right) on the natural frequency (mode 6) of the KW51 bridge.}
\label{fig:KW51_extended_model_interactions}
\end{figure}

As a second extension, we replace the additive effects from above by the term $f_{12}(z_{j1}(t),$ $z_{j2}(t))$ for the two covariates, allowing for an interacting effect on the system outputs. Figure~\ref{fig:KW51_extended_model_interactions} shows the (centered) two-dimensional functional intercept $\tilde{\alpha}(t, d_j)$ and the surface $f_{12}$ taking interactions of temperature and relative humidity into account in a smooth, nonparametric way.
As we can see, the strongest effect is along the temperature axis, but humidity is also relevant, with its strongest effect around 10°C. The functional intercept is flat, indicating no further daily or seasonal patterns. The yellow ``spike'' in the surface in the right plot of Figure~\ref{fig:KW51_extended_model_interactions} is because no data points with negative temperatures and high humidity are available. That is why $f_{12}$ cannot be fitted in that area, whereas data points with negative temperature and low humidity are found in the data set. For the latter, the highest natural frequency values (of mode 6) are predicted, hence the yellow spike. The overall model fit ($R^2 = 0.54$) is slightly better than with the additive model (compare Table~\ref{tab:R2_of_models}) and, as such, considerably better than the basic model. In summary, the temperature is the most influential covariate, and 
including it explicitly in the model provides a much better fit than using the two-dimensional intercept from Figure \ref{fig:KW51_reduced_model_estimated_effect} only, which leads to an $R^2$ of $= 0.25$. Please note that when calculating the $R^2$, we considered the fixed effects only. The predicted random effects are used for monitoring, as discussed below.  

\begin{table}[ht]
\centering
\caption{Summary of the $R^2$, overall number of observations, and number of profiles used in Phase I for the different compared models.}
\begin{tabular}{lrrrr}
    \toprule
    \tiny
    & \textbf{Basic} & \textbf{Output only} & \textbf{Additive} & \textbf{Interactions} \\ 
    \midrule
    $R^2$ & 0.43 & 0.25 & 0.52 & 0.54 \ \\ 
    \# observations & 3411 & 4488 & 2268 & 2268 \\ 
    \# profiles & 148 & 195 & 100 & 100 \\
    \bottomrule
\end{tabular}
\label{tab:R2_of_models}
\end{table}
%%%%%%%%%%%%%%%%%%%%%%%%%%%%%%%%%%%%%%%%%%%%%%%%%%%%%%%%%%%%%%%%%%
\subsubsection{Anomaly detection}\label{subsubsec:kw51_monitoring}
%%%%%%%%%%%%%%%%%%%%%%%%%%%%%%%%%%%%%%%%%%%%%%%%%%%%%%%%%%%%%%%%%%
Finally, we apply the second part of the CAFDA-SHM framework, the monitoring scheme, to the natural frequency data (mode 6) of the KW51 bridge. Two MEWMA chart configurations with different $\lambda$ are chosen and the retrofitting data is either included for online monitoring or omitted as in \cite{Maes.etal_2022}. As mentioned, Phase I consists of the first 200 days, but only 2268 observations or 100 profiles could be used for the extended model due to data availability, compare Table \ref{tab:R2_of_models}. The control charts are calibrated to an ARL$_0 = 370.4$ applying thresholds $h_4=16.25$ and $h_4=15.83$ for $\lambda=1$ and $\lambda=0.3$, respectively. Utilizing the estimated parameters of the best-performing model, the extended model with interactions (see Section~\ref{subsubsec:extended_model_results}) and the incoming data, the scores $\hat{\xi}_{r,g}$ from \eqref{eq:estSc2}, $r=4$, are used in \eqref{eq:MHD} to calculate the control statistic $T^2_g$ for day $g$ (compare Section~\ref{sec:ControlCharts}).

\begin{figure}[!htb]
\centering
\includegraphics[scale=.45]{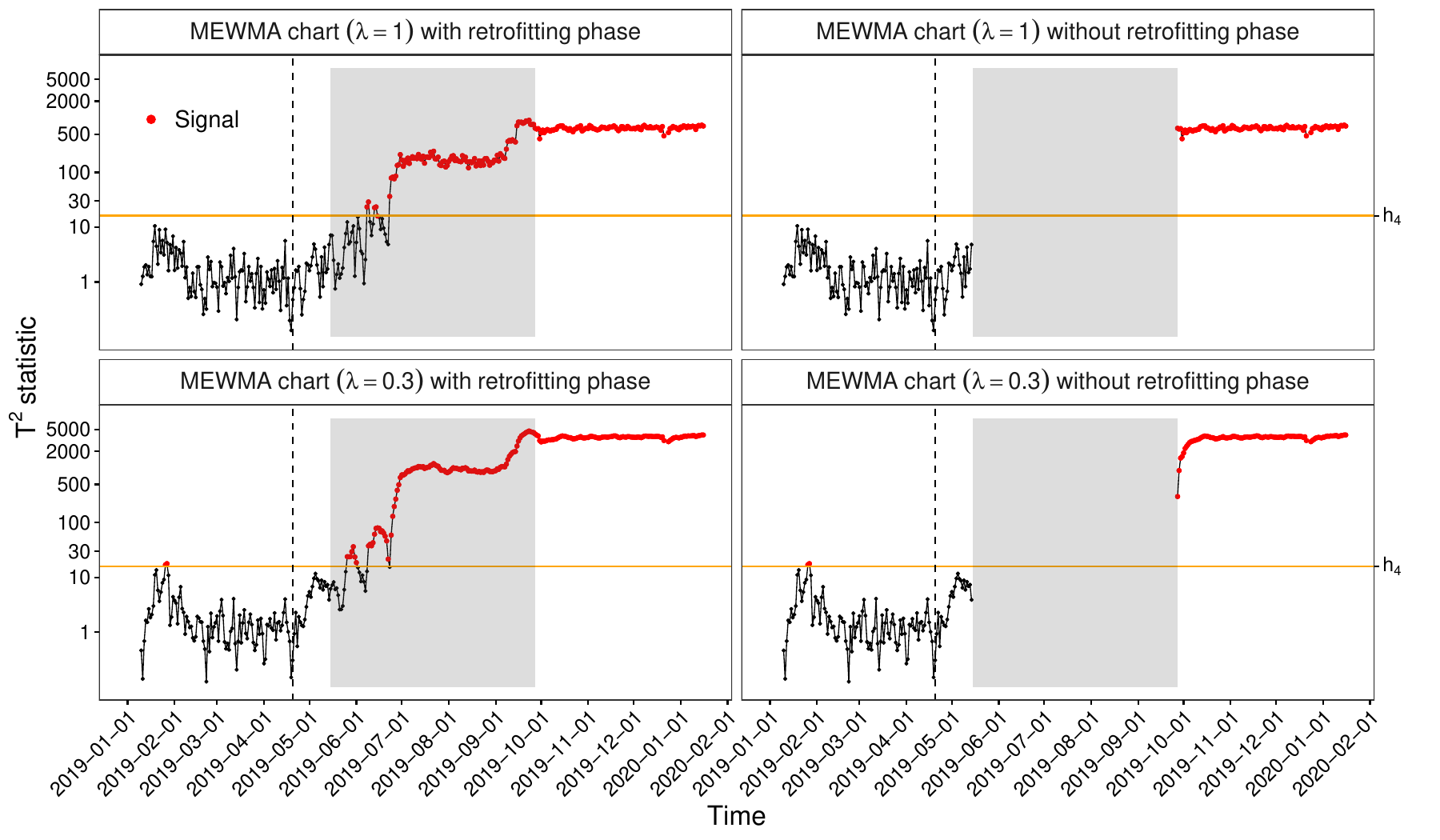}
\caption{MEWMA control charts for $\lambda=1$ (top row) and $\lambda=0.3$ (bottom row), on a logarithmic y-axis, using the extended functional model with interactions for the KW51 bridge with and without the retrofitting period (grey shaded area).}
\label{fig:MEWMA_chart_lambda_03_and1_extended}
\end{figure}

Figure \ref{fig:MEWMA_chart_lambda_03_and1_extended} displays the resulting control charts. In all MEWMA charts, online monitoring begins in Phase II after the dashed line, with the retrofitting period marked through the gray shaded area. For the control chart with $\lambda=1$ and consideration of the retrofitting phase (top left), the first signal was detected on June 8th, 2019, and for the chart without considering the retrofitting phase on October 28th, 2019, indicating a clear shift in the natural frequency of Mode 6. The control charts with $\lambda=0.3$ (bottom row) show a false alarm on January 26th and 27th, 2019. The bottom-left panel also reveals that compared to $\lambda=1$, an alarm was triggered on May 26, 2019, so eleven days earlier. However, comparing the different scenarios, it is noticeable that all four control charts issue several signals in a row, indicating a sustained change in the process within the retrofitting (gray shaded area) and after the retrofitting. Thus, employing our CAFDA framework, the control charts detect a change in the bridge's response caused by retrofitting. However, using the MEWMA, $\lambda<1$ leads to a faster detection than the Hotelling control chart.
In summary, the control charts used in our CAFDA framework can detect a retrofit-induced change in bridge behavior in the tracked natural frequencies, and faster detection is achieved through MEWMA parameterizations $\lambda<1$.
%%%%%%%%%%%%%%%%%%%%%%%%%%%%%%%%%%%%%%%%%%%%%%%%%%%%%%%%%%%%%%%%%%%%%%%%%%
\subsection{The Sachsengraben Viaduct}\label{sec:case_study_sachsengraben}
\begin{figure}[!ht]
\centering
\includegraphics[height=40mm]{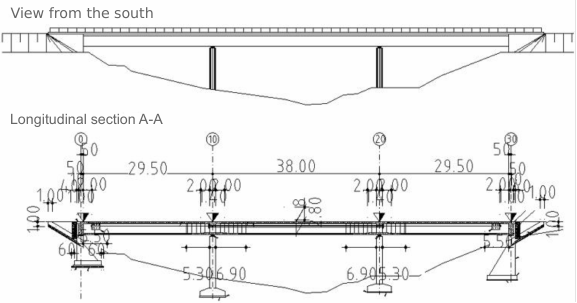}
\includegraphics[height=40mm]{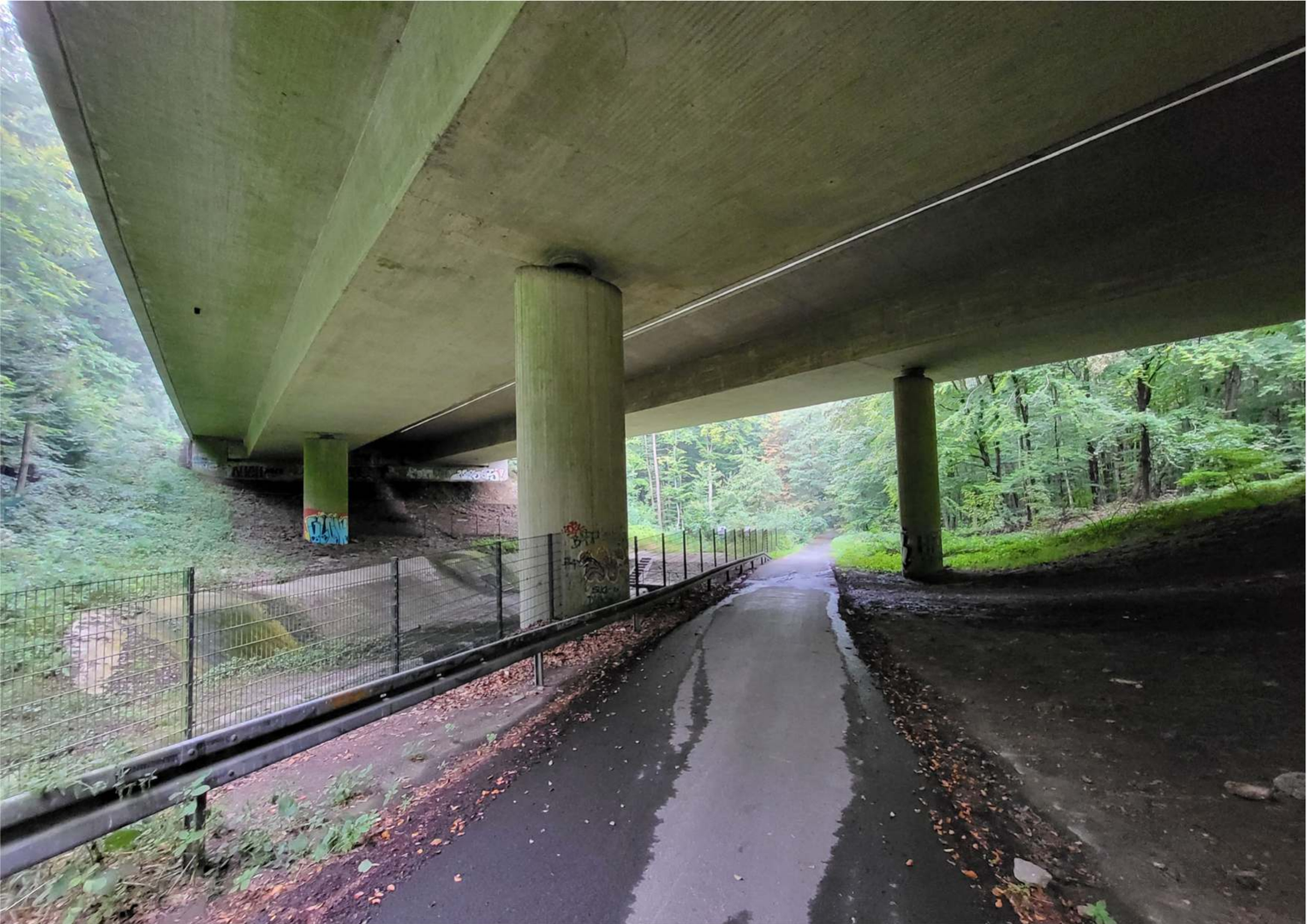}
\caption{Schematic representation (left) of the Sachsengraben viaduct from the OSIMAB report \citep{BASt_2021} and a photo of the bridge in 2023 (right).}
\label{fig:bridge_sachsengraben}
\end{figure}

%%%%%%%%%%%%%%%%%%%%%%%%%%%%%%%%%%%%%%%%%%%%%%%%%%%%%%%%%%%%%%%%%%%%%%%%%%
This section presents an application of the CAFDA-SHM framework on a second real-world SHM data set, the OSIMAB\footnote{Online-Sicherheits-Managementsystem für Brücken} data set \citep{BAST_2023}, which contains measurements from the Sachsengraben bridge, see Figure \ref{fig:bridge_sachsengraben}. The bridge is located on the A45 motorway in Germany and was constructed in 1971. It spans 98 meters and is made of prestressed concrete. The bridge's cross-section consists of a single-cell, box girder with a construction height of 2.8 meters. The webs, floor, and deck slab also have coves, and the end and support cross girders are stiffened. 
There are two separate substructures, and we will only focus on the northern one. The superstructure has three fields, divided by pillars. The two outer fields are 30 meters long, while the middle field is 38 meters long \citep{BASt_2021}.

Displacement was measured with eight displacement sensors per superstructure for 18 months, from January 1st, 2020, to August 1st, 2021, and with a sampling frequency of 100 Hz for the first 191 days and a forty-minute integration interval every hour. On day 191, the measurement switched to a 1 Hz sampling scheme, recording the data continuously. Ambient air temperature data recorded by a weather station and data from ten material temperature sensors with a sampling frequency of 1~Hz are also available. The entire data set was collected in the OSIMAB project. A sample of the recorded measurement data is publicly available \citep{BAST_2020}, and the complete data set is accessible via the 
 \citet{BAST_2023}.

\begin{figure}[!ht]
\centering
\includegraphics[scale=.4]{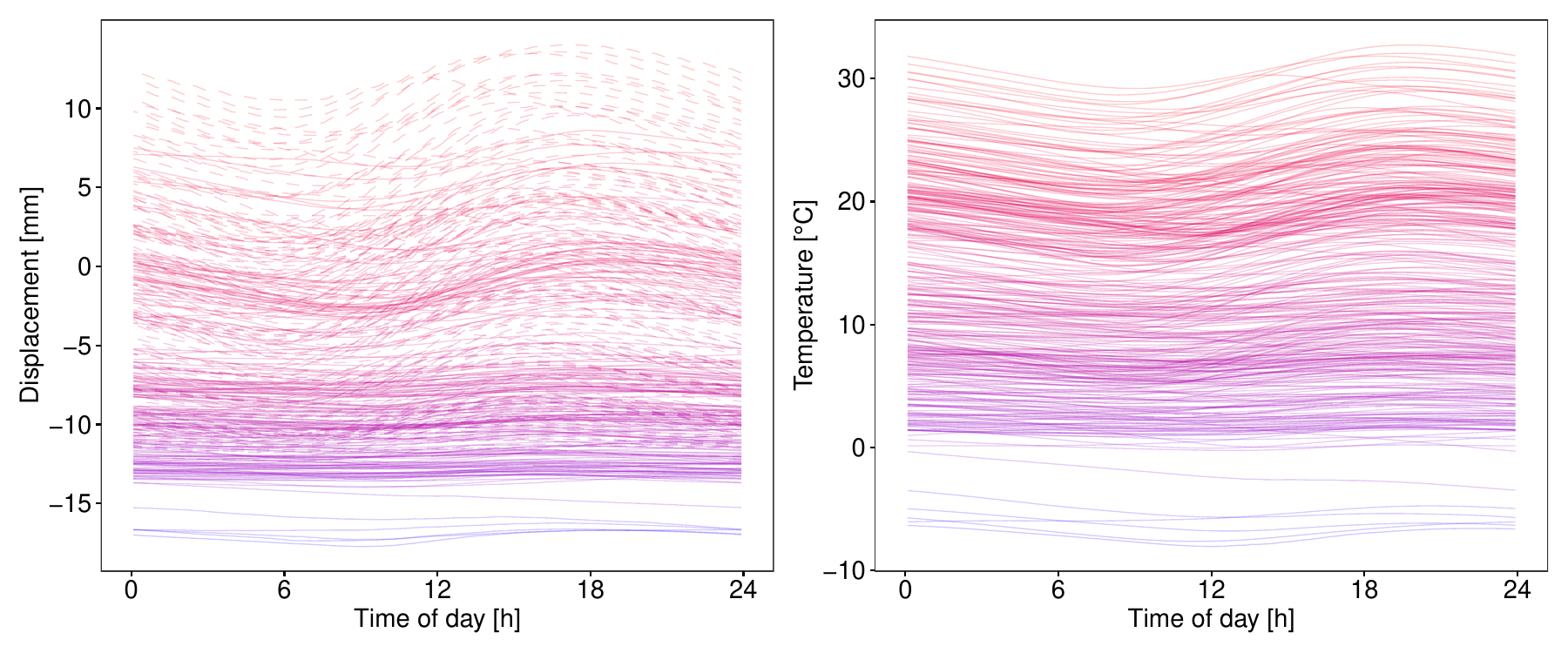}
\caption{Phase-I functional profiles of displacement sensor N\_F3\_WA\_NO (left panel) and temperature sensor N\_F1\_T\_1 (right panel) with a 10-minute sampling rate. The profiles are highlighted in color according to their average daily temperature.}
\label{fig:OSIMAB_profiles}
\end{figure}

Figure \ref{fig:OSIMAB_profiles} shows the Phase-I profiles of one displacement sensor (N\_F3\_WA\_NO, left) and measurements from the closest temperature sensor (N\_F1\_T\_1, right) for 369 days. The dashed lines represent the first 191 days, from February 10th, 2020, to August 19th, 2020. The displacement and temperature measurements were resampled from the higher frequencies (100~Hz and 1~Hz, respectively) by taking the median of the measurements within a 10-minute interval. This leads to a substantially larger number (144) of measurement points per day than with the KW51 bridge data, where the eigenfrequencies were obtained at 1-hour intervals. Furthermore, the 1.5-year measurement period means that data from each time of the year is available, and a two-dimensional intercept can be fitted that spans the entire year. The temperature values observed range from $-$10$^\circ$C to 30$^\circ$C. Out of 538 profiles in total, 287 displacement profiles and 461 temperature profiles did not exhibit any missing values, and overall, 20\% percent of the data is missing.\footnote{17 days (out of 538) have less than 13 observations per profile. Those days were excluded from model training.} This makes clear again that for SHM data, it is particularly important that the methods used are able to deal with missing data.

As mentioned above, the measurement recording process changed during Phase I. Nevertheless, we can still exploit the complete information available and capture a complete annual cycle in our framework. For doing so, we assume that changing the measurement systems' sampling rate does not impact the external influence of the covariates -- the fixed effects in our model. However, the functional random effects may show different behavior for the two time periods, which means that the eigenfunctions and variance components may differ. That is why two sets of residuals, before and after switching the sampling rate, are extracted to estimate two sets of eigenfunctions. For the first part of Phase I, five eigenfunctions, and for the second part, six eigenfunctions were sufficient to explain 99 \% of the variance. Both sets of eigenfunctions were then utilized in the refitted model in step 3~to estimate the fixed effects.

\begin{figure}[!htb]
\centering
\includegraphics[scale=.5]{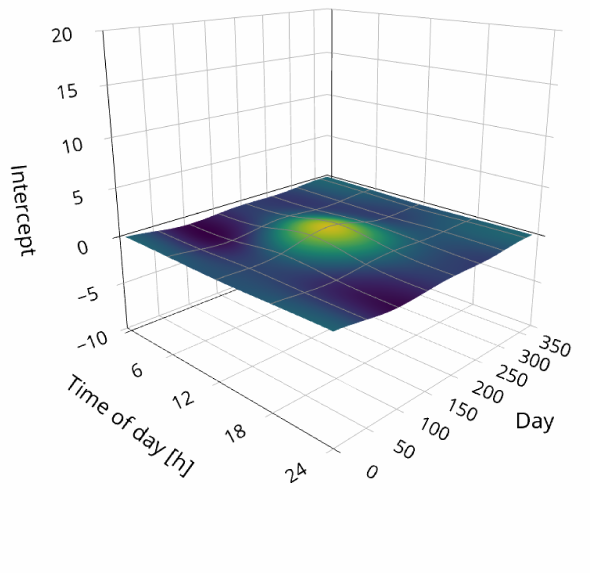}
\includegraphics[scale=.5]{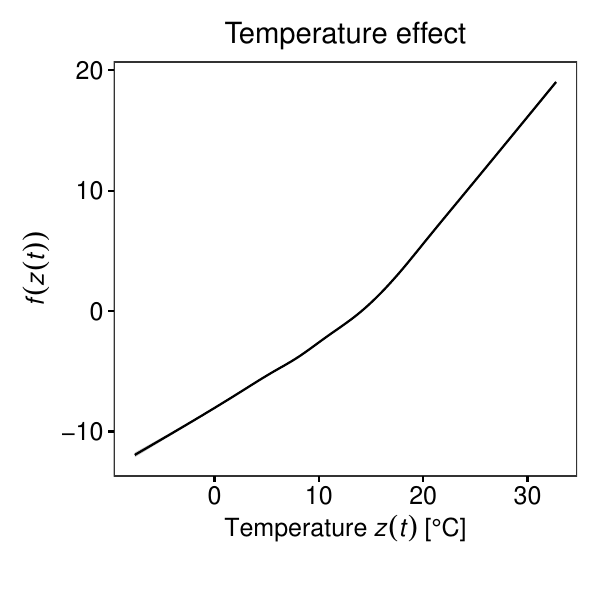}
\caption{Results of the extended functional modeling approach showing a two-dimensional functional intercept $\alpha(t, d{_j})$ (left) and the (non)-linear effect of temperature sensor N\_F1\_T\_1 (right) on the displacement sensor N\_F3\_WA\_NO.}
\label{fig:Osimab_extended_model_estimated_effects}
\end{figure}

Figure \ref{fig:Osimab_extended_model_estimated_effects} shows the resulting estimates of the fixed effects. On the left side, the two-dimensional functional intercept is rather flat over the entire year. Only a small bump during the summertime at around day 200 and in the middle of the day around noon is noticeable. The right panel in Figure \ref{fig:Osimab_extended_model_estimated_effects} shows an almost linear temperature effect with a tiny kink at around 15$^\circ$C. This linear covariate effect on the quasi-static response displacement is also found in the literature, compare with \cite{Han.etal_2021}. This strong temperature effect can already be seen from the colors in Figure~\ref{fig:OSIMAB_profiles}. However, it is worth mentioning that although the model accounts for the temperature effect, there is still variation captured by the two-dimensional intercept. Overall, the fixed effects model has an R$^2=0.99$.

\begin{figure}[!htb]
\centering
\includegraphics[scale=.5]{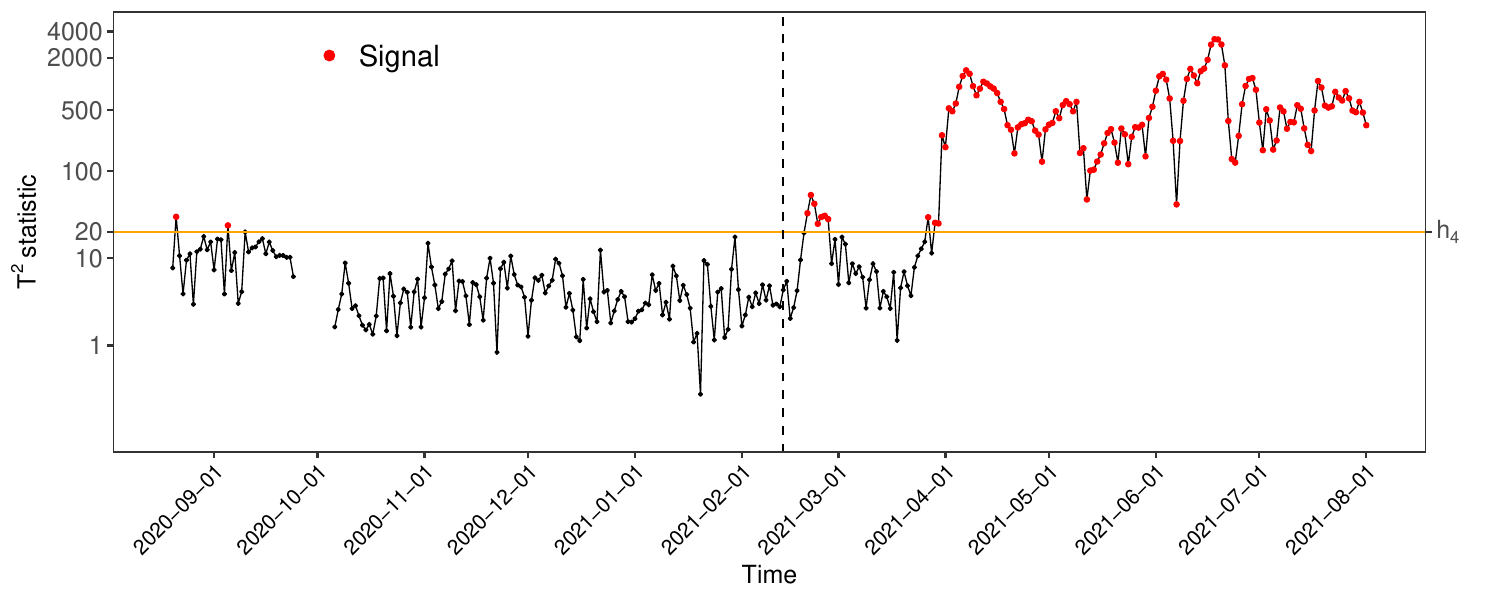}
\caption{MEWMA Control chart for $\lambda=1$ on a
logarithmic y-axis, using an extended functional model for the Sachsengraben viaduct.}
\label{fig:Osimab_extended_model_control_chart}
\end{figure}

Similar to applying the control chart for anomaly detection in Section \ref{subsubsec:kw51_monitoring} and as shown in Figure~\ref{fig:flowchart_procedure}, we set up the control chart, calibrating it to an ARL$_0 = 370.4$. For monitoring purposes, only the second set of eigenfunctions and variance components are used, as those were estimated on the reconfigured measurement system with the new sampling rate. Figure \ref{fig:Osimab_extended_model_control_chart} displays the control chart for $\lambda=1$ and a chart threshold $h_4=20.06$. It is visible that besides the two false alarms in January 2021 in the Phase-I data, the measurements before the dashed vertical line seem to be in control, and the sensor is working as expected. The dashed line marks the beginning of the online monitoring on February 13th, 2021. \cite{BASt_2021} reported that a sensor malfunction induced by a rust film blocking the sensor was noticed and confirmed at the end of March. The control chart in Figure \ref{fig:Osimab_extended_model_control_chart} detects this anomaly. However, it additionally detects some irregularities at the end of February. Presumably, sensor malfunction already started before the end of March and this was detected by the control chart. 

In summary, our real-world data analyses showed that the CAFDA-SHM framework can be used for the covariate adjustment and monitoring of different types of response variables: dynamic responses, as shown in Section~\ref{sec:KW51}, and quasi-static responses, as demonstrated in Section~\ref{sec:case_study_sachsengraben}. 

\section{Concluding remarks}\label{sec:Conclusion}
The modeling and monitoring methodology presented in this paper combines several parts of a structural health monitoring system in one unified framework (CAFDA-SHM). It is of high practical relevance, as it tackles important challenges in SHM. The discussed functional data approach offers the flexibility to model recurring daily and yearly patterns alongside environmental or operational influences in a highly adaptable and interpretable semiparametric manner. It can handle missing observations and effectively accounts for variations and correlations in the error process through functional principal components. Furthermore, FPCA extracts interpretable, data-based features, which are then utilized for monitoring within a sound statistical framework, thus providing a reliable basis for decision-making. It is hence planned to implement and test the methods presented in this article as part of a larger online SHM system~\citep{Kessler.etal_2024}.

Although the proposed modeling approach is an input-output method, it contains an output-only FPCA-based version as a particular case. The latter has advantages over existing approaches as it accounts for daily and yearly patterns typically ignored by common PCA-based methods. Utilizing state-of-the-art functional additive mixed models enables nonlinear modeling, as often needed in SHM. Our approach is flexible in handling diverse data types, from sparse and aggregated to high-resolution, dense data. Although the focus of this article was on concurrent models, the framework of functional additive mixed models includes various other options, such as historical functional effects or the common linear function-on-function approach (compare Appendix~\ref{sec:model_extensions}). 

Finally, there are three noteworthy limitations. First, if the error process becomes close to white noise, e.g., because the covariates' explanatory power is very high, the functional approach no longer provides much of an advantage over a standard, non-functional response surface model. 
Second, in the functional additve mixed model we implicitly assume that the error process and the functional random effets are (at least approximately) Gaussian. However, if outliers are present in the data~\citep[compare, e.g.,][]{Capezza.etal_2024}, an additional step to identify these outliers would need to be included in a preprocessing step to our framework. Third, the framework presented in this paper is designed for univariate functional responses. Future research will involve modeling multiple system outputs simultaneously, such as natural frequencies of multiple modes.

%%%%%%%%%%%%%%%%%%%%%%%%%%%%
\section*{Funding Statement}
%%%%%%%%%%%%%%%%%%%%%%%%%%%%
This research is funded by dtec.bw -- Digitalization and Technology Research Center of the Bundeswehr. dtec.bw is funded by the European Union -- NextGenerationEU. We thank the hpc.bw team for their support within the joint dtec.bw project ``HPC for semi-parametric statistical modeling on massive datasets''.
%%%%%%%%%%%%%%%%%%%%%%%%%%%%%%%%%%%%%%
\section*{Data Availability Statement}
%%%%%%%%%%%%%%%%%%%%%%%%%%%%%%%%%%%%%%
The vibration, temperature, and relative humidity data for the railway bridge KW51 are available from \cite{Maes.Lombaert_2020}. The strain and temperature data for the viaduct Sachsengraben are available from \cite{BAST_2023}. 
The source code for reproducing the main results is available at \href{https://github.com/wittenberg/CAFDA-SHM_code}{https://github.com/wittenberg/CAFDA-SHM\_code}.

%%%%%%%%%%%%%%%%%%%
\section*{Software}
%%%%%%%%%%%%%%%%%%%
The main analysis used the statistical software \texttt{R} \citep{R_2023}. In addition, functions from the following R packages were used. Some functions from the \texttt{funData} \citep{Happ-Kurz_2020} were used to generate the artificial data. The functional additive modeling, estimation, and mixed modeling were carried out with \texttt{mgcv} \citep{Wood_2017}. Functional principal component analysis utilized \texttt{refund} \citep{Goldsmith.etal_2022}. The control limits of the MEWMA control chart were calculated using \texttt{spc} R package \citep{Knoth_2022}.
%%%%%%%%%%%%%%%%%%%%%%%%%%%%
%%%%%%%%%%%%%%%%%%%

%%%%%%%%%%%%%%%%%%%
\section*{Appendix}
\begin{appendix}
%%%%%%%%%%%%%%%%%%%
%%%%%%%%%%%%%%%%%%%%%%%%%%%%%%%%%%%%%%%%%%%%%%%%%%%%%%%%%%%%%%
\section{Further modeling options}\label{sec:model_extensions}
%%%%%%%%%%%%%%%%%%%%%%%%%%%%%%%%%%%%%%%%%%%%%%%%%%%%%%%%%%%%%%
This Section provides an overview of various modeling options beyond the models discussed in the main paper. All models considered in Sections \ref{sec:FDA}, \ref{sec:Simulation}, and \ref{sec:KW51} and \ref{sec:case_study_sachsengraben} were so-called \emph{concurrent} models, where it is assumed that the system output at time $t$ is influenced by covariate $z$ measured at the same time $t$ only. This makes sense for covariates such as the temperature of the structure itself, as available for the KW51 bridge and the Sachsengraben viaduct. However, if only ambient temperature is given, it seems more reasonable to assume that the temperature over the recent past, let us say three hours, is relevant. Appropriate models can be specified through so-called \emph{historical} (functional) effects.
\begin{itemize}
    \item In the simplest case of a \textbf{linear effect} that does not change over the day, we have
$$u_j(t) = \ldots + \int_0^h \beta(s)z_j(t-s)ds + \ldots \, ,$$
where $h$ denotes the time limit to look into the past, e.g., three hours. Typically, it is recommended to choose a rather large value here because, if $z$ in that time region is not relevant for $u$ anymore, $\beta(s)$ will be fit to tend towards zero for $s \rightarrow h$ (given there is enough data available to learn from). 
\item If the effect of $z$ is allowed to \textbf{change over the course of the day} $\int_0^h \beta(s)z_j(t-s)ds$ turns into $\int_0^h \beta(s,t)z_j(t-s)ds$.
\item In the \textbf{non-linear setting}, we have $\int_0^h f(z_j(t-s),s)ds$ if the (historical) effect is constant across $t$ (i.e., the course of the day) and $\int_0^h f(z_j(t-s),s,t)ds$ otherwise.
\end{itemize}
All those models fit into the framework of functional additive mixed models as well. Instead of the historical functional effects, we could also write 
$$u_j(t) = \ldots + \int_\mathcal{T} \beta(s,t)z_j(s)ds + \ldots \, ,$$
which is typically denoted as ``linear function-on-function regression''. In fact, this is the most popular model for function-on-function regression, and, e.g., used by \cite{Centofanti.etal_2021}. The corresponding non-linear/smooth version would be
$$u_j(t) = \ldots + \int_\mathcal{T} f(z_j(s),s,t)ds + \ldots \, .$$
However, both models do not seem sensible approaches for SHM data as considered here because ``future'' observations $z_j(s)$, with $s > t$, would be allowed to affect ``present'' $u_j(t)$.
\end{appendix}
%%%%%%%%%%%%%%

\begin{thebibliography}{67}
\providecommand{\natexlab}[1]{#1}
\providecommand{\url}[1]{\texttt{#1}}
\expandafter\ifx\csname urlstyle\endcsname\relax
  \providecommand{\doi}[1]{doi: #1}\else
  \providecommand{\doi}{doi: \begingroup \urlstyle{rm}\Url}\fi

\bibitem[Abramowitz and Stegun(1964)]{Abramowitz.Setgun_1964}
M.~Abramowitz and I.~A. Stegun.
\newblock \emph{Handbook of mathematical functions: with formulas, graphs, and
  mathematical tables}.
\newblock Dover Publications, New York, 1964.

\bibitem[Anastasopoulos et~al.(2021)Anastasopoulos, {De Roeck}, and
  Reynders]{Anastasopoulos.etal_2021}
D.~Anastasopoulos, G.~{De Roeck}, and E.~P. Reynders.
\newblock One-year operational modal analysis of a steel bridge from
  high-resolution macrostrain monitoring: Influence of temperature vs.
  retrofitting.
\newblock \emph{Mech Syst Signal Process}, 161:\penalty0 107951, 2021.
\newblock \doi{10.1016/j.ymssp.2021.107951}.

\bibitem[Avci et~al.(2021)Avci, Abdeljaber, Kiranyaz, Hussein, Gabbouj, and
  Inman]{Avci.etal_2021}
O.~Avci, O.~Abdeljaber, S.~Kiranyaz, M.~Hussein, M.~Gabbouj, and D.~J. Inman.
\newblock {A review of vibration-based damage detection in civil structures:
  From traditional methods to Machine Learning and Deep Learning applications}.
\newblock \emph{Mech Syst Signal Process}, 147:\penalty0 107077, 2021.
\newblock \doi{10.1016/j.ymssp.2020.107077}.

\bibitem[{Bundesanstalt für Straßenwesen (BASt)}(2020)]{BAST_2020}
{Bundesanstalt für Straßenwesen (BASt)}.
\newblock \emph{{OSIMAB} – {O}nline-{S}icherheits-{M}anagementsystem f\"ur
  {B}r\"ucken}, 2020.
\newblock Data licence Germany – attribution – Version 2.0.

\bibitem[{Bundesanstalt für Straßenwesen (BASt)}(2021)]{BASt_2021}
{Bundesanstalt für Straßenwesen (BASt)}.
\newblock Verbundprojekt {OSIMAB} ({O}nline {S}icherheits-{M}angementssystem
  für {B}rücken) : Gesamtabschlussbericht, 2021.
\newblock URL \url{https://www.tib.eu/de/suchen/id/TIBKAT\%3A1787930734}.

\bibitem[{Bundesanstalt für Straßenwesen (BASt)}(2023)]{BAST_2023}
{Bundesanstalt für Straßenwesen (BASt)}.
\newblock {OSIMAB} – {O}nline-{S}icherheits-{M}anagementsystem f\"ur
  {B}r\"ucken, 2023.
\newblock URL
  \url{https://www.bast.de/DE/Publikationen/Daten/Brueckenbau/Downloads/OSIMAB.html}.
\newblock Online; accessed 15-October-2023.

\bibitem[Capezza et~al.(2023)Capezza, Centofanti, Lepore, Menafoglio, Palumbo,
  and Vantini]{Capezza.etal_2023c}
C.~Capezza, F.~Centofanti, A.~Lepore, A.~Menafoglio, B.~Palumbo, and
  S.~Vantini.
\newblock funcharts: control charts for multivariate functional data in {R}.
\newblock \emph{J Qual Technol}, 55\penalty0 (5):\penalty0 566--583, 2023.
\newblock \doi{10.1080/00224065.2023.2219012}.

\bibitem[Capezza et~al.(2024)Capezza, Centofanti, Lepore, and
  Palumbo]{Capezza.etal_2024}
C.~Capezza, F.~Centofanti, A.~Lepore, and B.~Palumbo.
\newblock {Robust Multivariate Functional Control Chart}.
\newblock \emph{Technometrics}, 66\penalty0 (4):\penalty0 531--547, 2024.
\newblock \doi{10.1080/00401706.2024.2327346}.

\bibitem[Centofanti et~al.(2021)Centofanti, Lepore, Menafoglio, Palumbo, and
  Vantini]{Centofanti.etal_2021}
F.~Centofanti, A.~Lepore, A.~Menafoglio, B.~Palumbo, and S.~Vantini.
\newblock {Functional Regression Control Chart}.
\newblock \emph{Technometrics}, 63\penalty0 (3):\penalty0 281--294, 2021.
\newblock \doi{10.1080/00401706.2020.1753581}.

\bibitem[Chen et~al.(2018)Chen, Bao, Li, and Spencer]{Chen.etal_2018}
Z.~Chen, Y.~Bao, H.~Li, and B.~F. Spencer.
\newblock A novel distribution regression approach for data loss compensation
  in structural health monitoring.
\newblock \emph{Struct Health Monit}, 17:\penalty0 1473--1490, 2018.
\newblock \doi{10.1177/1475921717745719}.

\bibitem[Chen et~al.(2019)Chen, Bao, Li, and Spencer]{Chen.etal_2019}
Z.~Chen, Y.~Bao, H.~Li, and B.~F. Spencer.
\newblock {LQD-RKHS}-based distribution-to-distribution regression methodology
  for restoring the probability distributions of missing {SHM} data.
\newblock \emph{Mech Syst Signal Process}, 121:\penalty0 655--674, 2019.
\newblock \doi{10.1016/j.ymssp.2018.11.052}.

\bibitem[Chen et~al.(2020)Chen, Bao, Tang, Chen, and Li]{Chen.etal_2020}
Z.~Chen, Y.~Bao, Z.~Tang, J.~Chen, and H.~Li.
\newblock Clarifying and quantifying the geometric correlation for probability
  distributions of inter-sensor monitoring data: A functional data analytic
  methodology.
\newblock \emph{Mech Syst Signal Process}, 138:\penalty0 106540, 2020.
\newblock \doi{10.1016/j.ymssp.2019.106540}.

\bibitem[Chen et~al.(2021)Chen, Lei, Bao, Deng, Zhang, and Li]{Chen.etal_2021b}
Z.~Chen, X.~Lei, Y.~Bao, F.~Deng, Y.~Zhang, and H.~Li.
\newblock Uncertainty quantification for the distribution-to-warping function
  regression method used in distribution reconstruction of missing structural
  health monitoring data.
\newblock \emph{Struct Health Monit}, 20\penalty0 (6):\penalty0 3436--3452,
  2021.
\newblock \doi{10.1177/1475921721993381}.

\bibitem[Comanducci et~al.(2016)Comanducci, Magalh{\~{a}}es, Ubertini, and
  Cunha]{Comanducci.etal_2016}
G.~Comanducci, F.~Magalh{\~{a}}es, F.~Ubertini, and {\'{A}}.~Cunha.
\newblock On vibration-based damage detection by multivariate statistical
  techniques: {A}pplication to a long-span arch bridge.
\newblock \emph{Struct Health Monit}, 15\penalty0 (5):\penalty0 505--524, 2016.
\newblock \doi{10.1177/1475921716650630}.

\bibitem[Cross et~al.(2013)Cross, Koo, Brownjohn, and Worden]{Cross.etal_2013}
E.~Cross, K.~Koo, J.~Brownjohn, and K.~Worden.
\newblock Long-term monitoring and data analysis of the {T}amar {B}ridge.
\newblock \emph{Mech Syst Signal Process}, 35\penalty0 (1):\penalty0 16--34,
  2013.
\newblock \doi{10.1016/j.ymssp.2012.08.026}.

\bibitem[Cross et~al.(2012)Cross, Manson, Worden, and Pierce]{Cross.etal_2012}
E.~J. Cross, G.~Manson, K.~Worden, and S.~G. Pierce.
\newblock Features for damage detection with insensitivity to environmental and
  operational variations.
\newblock \emph{Proc R Soc A}, 468\penalty0 (2148):\penalty0 4098--4122, 2012.
\newblock \doi{10.1098/rspa.2012.0031}.

\bibitem[de~Boor(1978)]{deBoor_1978}
C.~de~Boor.
\newblock \emph{{A Practical Guide to Splines}}.
\newblock Springer-Verlag,, New York, 1978.

\bibitem[Deraemaeker et~al.(2008)Deraemaeker, Reynders, Roeck, and
  Kullaa]{Deraemaeker.etal_2008}
A.~Deraemaeker, E.~Reynders, G.~D. Roeck, and J.~Kullaa.
\newblock Vibration-based structural health monitoring using output-only
  measurements under changing environment.
\newblock \emph{Mech Syst Signal Process}, 22\penalty0 (1):\penalty0 34--56,
  2008.
\newblock \doi{10.1016/j.ymssp.2007.07.004}.

\bibitem[Dierckx(1993)]{Dierckx_1993}
P.~Dierckx.
\newblock \emph{{Curve and Surface Fitting with Splines}}.
\newblock Oxford University Press, New York, 1993.

\bibitem[Eilers and Marx(1996)]{Eilers.Marx_1996}
P.~H.~C. Eilers and B.~D. Marx.
\newblock {Flexible Smoothing with B-splines and Penalties}.
\newblock \emph{Stat Sci}, 11\penalty0 (2):\penalty0 89--121, 1996.
\newblock \doi{10.1214/ss/1038425655}.

\bibitem[Gertheiss et~al.(2017)Gertheiss, Goldsmith, and
  Staicu]{Gertheiss.etal_2017}
J.~Gertheiss, J.~Goldsmith, and A.-M. Staicu.
\newblock A note on modeling sparse exponential-family functional response
  curves.
\newblock \emph{Comput Stat Data Anal}, 105:\penalty0 46 -- 52, 2017.
\newblock \doi{10.1016/j.csda.2016.07.010}.

\bibitem[Gertheiss et~al.(2024)Gertheiss, Rügamer, Liew, and
  Greven]{Gertheiss.etal_2024}
J.~Gertheiss, D.~Rügamer, B.~X.~W. Liew, and S.~Greven.
\newblock {Functional Data Analysis: An Introduction and Recent Developments}.
\newblock \emph{Biom J}, 66\penalty0 (7):\penalty0 e202300363, 2024.
\newblock \doi{10.1002/bimj.202300363}.

\bibitem[Goldsmith et~al.(2022)Goldsmith, Scheipl, Huang, Wrobel, Di, Gellar,
  Harezlak, McLean, Swihart, Xiao, Crainiceanu, and Reiss]{Goldsmith.etal_2022}
J.~Goldsmith, F.~Scheipl, L.~Huang, J.~Wrobel, C.~Di, J.~Gellar, J.~Harezlak,
  M.~W. McLean, B.~Swihart, L.~Xiao, C.~Crainiceanu, and P.~T. Reiss.
\newblock \emph{{refund: Regression with Functional Data}}, 2022.
\newblock URL \url{https://CRAN.R-project.org/package=refund}.
\newblock R package version 0.1-26.

\bibitem[Greven and Scheipl(2017)]{Greven.Scheipl_2017}
S.~Greven and F.~Scheipl.
\newblock A general framework for functional regression modelling.
\newblock \emph{Stat Modelling}, 17\penalty0 (1-2):\penalty0 1--35, 2017.
\newblock \doi{10.1177/1471082X16681317}.

\bibitem[Han et~al.(2021)Han, Ma, Xu, and Liu]{Han.etal_2021}
Q.~Han, Q.~Ma, J.~Xu, and M.~Liu.
\newblock Structural health monitoring research under varying temperature
  condition: a review.
\newblock \emph{J Civ Struct Health Monit}, 11:\penalty0 149--173, 2021.
\newblock \doi{10.1007/s13349-020-00444-x}.

\bibitem[Happ-Kurz(2020)]{Happ-Kurz_2020}
C.~Happ-Kurz.
\newblock {O}bject-{O}riented {S}oftware for {F}unctional {D}ata.
\newblock \emph{J Stat Softw}, 93\penalty0 (5):\penalty0 1--38, 2020.
\newblock \doi{10.18637/jss.v093.i05}.

\bibitem[Hou and Xia(2021)]{Hou.Xia_2021}
R.~Hou and Y.~Xia.
\newblock Review on the new development of vibration-based damage
  identification for civil engineering structures: 2010–2019.
\newblock \emph{J Sound Vib}, 491:\penalty0 115741, 2021.
\newblock \doi{10.1016/j.jsv.2020.115741}.

\bibitem[Hu et~al.(2012)Hu, Moutinho, Caetano, Magalhães, and Álvaro
  Cunha]{Hu.etal_2012}
W.-H. Hu, C.~Moutinho, E.~Caetano, F.~Magalhães, and Álvaro Cunha.
\newblock Continuous dynamic monitoring of a lively footbridge for
  serviceability assessment and damage detection.
\newblock \emph{Mech Syst Signal Process}, 33:\penalty0 38--55, 2012.
\newblock \doi{10.1016/j.ymssp.2012.05.012}.

\bibitem[Hunter(1986)]{Hunter_1986}
J.~S. Hunter.
\newblock {The Exponentially Weighted Moving Average}.
\newblock \emph{J Qual Technol}, 18\penalty0 (4):\penalty0 203--210, 1986.
\newblock \doi{10.1080/00224065.1986.11979014}.

\bibitem[Jiang et~al.(2021)Jiang, Wan, Yang, Ding, and Xue]{Jiang.etal_2021}
H.~Jiang, C.~Wan, K.~Yang, Y.~Ding, and S.~Xue.
\newblock Modeling relationships for field strain data under thermal effects
  using functional data analysis.
\newblock \emph{Measurement}, 177:\penalty0 109279, 2021.
\newblock \doi{10.1016/j.measurement.2021.109279}.

\bibitem[Karhunen(1947)]{Karhunen_1947}
K.~Karhunen.
\newblock {Ü}ber lineare {M}ethoden in der {W}ahrscheinlichkeitsrechnung.
\newblock \emph{Ann Acad Sci Fennicae Ser A I Math-Phys}, 37:\penalty0 1--79,
  1947.

\bibitem[Kessler et~al.(2024)Kessler, Mendler, Klemm, Jaelani, Köhncke, and
  Gündel]{Kessler.etal_2024}
S.~Kessler, A.~Mendler, A.~Klemm, Y.~Jaelani, M.~Köhncke, and M.~Gündel.
\newblock {A highway bridge—A database for digital methods}.
\newblock \emph{Struct Concr}, 25\penalty0 (5):\penalty0 3046--3049, 2024.
\newblock \doi{10.1002/suco.202400119}.

\bibitem[Knoth(2017)]{Knoth_2017}
S.~Knoth.
\newblock {ARL} numerics for {MEWMA} {C}harts.
\newblock \emph{J Qual Technol}, 49\penalty0 (1):\penalty0 78--89, 2017.
\newblock \doi{10.1080/00224065.2017.11918186}.

\bibitem[Knoth(2022)]{Knoth_2022}
S.~Knoth.
\newblock \emph{spc: Statistical Process Control -- Calculation of ARL and
  Other Control Chart Performance Measures}, 2022.
\newblock URL \url{https://CRAN.R-project.org/package=spc}.
\newblock R package version 0.6.7.

\bibitem[Knoth and Schmid(2004)]{Knoth.Schmid_2004}
S.~Knoth and W.~Schmid.
\newblock {Control Charts for Time Series: A Review}.
\newblock In H.-J. Lenz and P.-T. Wilrich, editors, \emph{Frontiers in
  Statistical Quality Control 7}, pages 210--236, Heidelberg, 2004.
  Physica-Verlag HD.

\bibitem[Kullaa(2011)]{Kullaa_2011}
J.~Kullaa.
\newblock Distinguishing between sensor fault, structural damage, and
  environmental or operational effects in structural health monitoring.
\newblock \emph{Mech Syst Signal Process}, 25\penalty0 (8):\penalty0
  2976--2989, 2011.
\newblock \doi{10.1016/j.ymssp.2011.05.017}.

\bibitem[Lei et~al.(2023{\natexlab{a}})Lei, Chen, and Li]{Lei.etal_2023b}
X.~Lei, Z.~Chen, and H.~Li.
\newblock {Functional Outlier Detection for Density-Valued Data with
  Application to Robustify Distribution-to-Distribution Regression}.
\newblock \emph{Technometrics}, 65\penalty0 (3):\penalty0 351--362,
  2023{\natexlab{a}}.
\newblock \doi{10.1080/00401706.2022.2164063}.

\bibitem[Lei et~al.(2023{\natexlab{b}})Lei, Chen, Li, and Wei]{Lei.etal_2023}
X.~Lei, Z.~Chen, H.~Li, and S.~Wei.
\newblock Detecting and testing multiple change points in distributions of
  damage-sensitive feature data for data-driven structural condition
  assessment: A distributional time series change-point analytic approach.
\newblock \emph{Mech Syst Signal Process}, 196:\penalty0 110344,
  2023{\natexlab{b}}.
\newblock \doi{10.1016/j.ymssp.2023.110344}.

\bibitem[Lei et~al.(2023{\natexlab{c}})Lei, Chen, Li, and Wei]{Lei.etal_2023c}
X.~Lei, Z.~Chen, H.~Li, and S.~Wei.
\newblock {A change-point detection method for detecting and locating the
  abrupt changes in distributions of damage-sensitive features of SHM data,
  with application to structural condition assessment}.
\newblock \emph{Struct Health Monit}, 22\penalty0 (2):\penalty0 1161--1179,
  2023{\natexlab{c}}.
\newblock \doi{10.1177/14759217221101320}.

\bibitem[Lowry et~al.(1992)Lowry, Woodall, Champ, and Rigdon]{Lowry.etal_1992}
C.~A. Lowry, W.~H. Woodall, C.~W. Champ, and S.~E. Rigdon.
\newblock {A Multivariate Exponentially Weighted Moving Average Control Chart}.
\newblock \emph{Technometrics}, 34\penalty0 (1):\penalty0 46--53, 1992.
\newblock \doi{10.2307/1269551}.

\bibitem[Loève(1946)]{Loeve_1946}
M.~Loève.
\newblock Fonctions aléatoires du second ordre.
\newblock \emph{La Revue Scientifique}, 84:\penalty0 195--206, 1946.

\bibitem[Maes and Lombaert(2020)]{Maes.Lombaert_2020}
K.~Maes and G.~Lombaert.
\newblock {Monitoring Railway Bridge {KW51} Before, During, and After
  Retrofitting. V1.0}, 2020.

\bibitem[Maes and Lombaert(2021)]{Maes.Lombaert_2021}
K.~Maes and G.~Lombaert.
\newblock {Monitoring Railway Bridge KW51 Before, During, and After
  Retrofitting}.
\newblock \emph{J Bridge Eng}, 26\penalty0 (3):\penalty0 04721001, 2021.
\newblock \doi{10.1061/(ASCE)BE.1943-5592.0001668}.

\bibitem[Maes et~al.(2022)Maes, {Van Meerbeeck}, Reynders, and
  Lombaert]{Maes.etal_2022}
K.~Maes, L.~{Van Meerbeeck}, E.~Reynders, and G.~Lombaert.
\newblock {Validation of vibration-based structural health monitoring on
  retrofitted railway bridge KW51}.
\newblock \emph{Mech Syst Signal Process}, 165:\penalty0 108380, 2022.
\newblock \doi{10.1016/j.ymssp.2021.108380}.

\bibitem[Magalh{\~{a}}es et~al.(2012)Magalh{\~{a}}es, Cunha, and
  Caetano]{Magalhaes.etal_2012}
F.~Magalh{\~{a}}es, A.~Cunha, and E.~Caetano.
\newblock {Vibration based structural health monitoring of an arch bridge: From
  automated {OMA} to damage detection}.
\newblock \emph{Mech Syst Signal Process}, 28:\penalty0 212--228, 2012.
\newblock \doi{10.1016/j.ymssp.2011.06.011}.

\bibitem[Maleki et~al.(2018)Maleki, Amiri, and Castagliola]{Maleki.etal_2018}
M.~R. Maleki, A.~Amiri, and P.~Castagliola.
\newblock An overview on recent profile monitoring papers (2008–2018) based
  on conceptual classification scheme.
\newblock \emph{Comput Ind Eng}, 126:\penalty0 705--728, 2018.
\newblock \doi{10.1016/j.cie.2018.10.008}.

\bibitem[Mercer(1909)]{Mercer_1909}
J.~Mercer.
\newblock Functions of positive and negative type, and their connection with
  the theory of integral equations.
\newblock \emph{Proceedings of the Royal Society of London. Series A,
  Containing Papers of a Mathematical and Physical Character}, 83\penalty0
  (559):\penalty0 69--70, 1909.
\newblock \doi{10.1098/rspa.1909.0075}.

\bibitem[Momeni and Ebrahimkhanlou(2022)]{Momeni.Ebrahimkhanlou_2022}
H.~Momeni and A.~Ebrahimkhanlou.
\newblock High-dimensional data analytics in structural health monitoring and
  non-destructive evaluation: a review paper.
\newblock \emph{Smart Mater Struct}, 31\penalty0 (4):\penalty0 043001, 2022.
\newblock \doi{10.1088/1361-665X/ac50f4}.

\bibitem[Moser and Moaveni(2011)]{Moser.Moaveni_2011}
P.~Moser and B.~Moaveni.
\newblock {Environmental effects on the identified natural frequencies of the
  Dowling Hall Footbridge}.
\newblock \emph{Mech Syst Signal Process}, 25\penalty0 (7):\penalty0
  2336--2357, 2011.
\newblock \doi{10.1016/j.ymssp.2011.03.005}.

\bibitem[{R Core Team}(2023)]{R_2023}
{R Core Team}.
\newblock \emph{{R: A Language and Environment for Statistical Computing}}.
\newblock R Foundation for Statistical Computing, Vienna, Austria, 2023.
\newblock URL \url{https://www.R-project.org/}.

\bibitem[Ramsay and Silverman(2005)]{Ramsay.Silvermann_2005}
J.~O. Ramsay and B.~W. Silverman.
\newblock \emph{Functional Data Analysis}.
\newblock Springer New York, 2005.
\newblock \doi{10.1007/b98888}.

\bibitem[Rice and Silverman(1991)]{Rice.Silvermann_1991}
J.~A. Rice and B.~W. Silverman.
\newblock {Estimating the Mean and Covariance Structure Nonparametrically When
  the Data are Curves}.
\newblock \emph{J R Stat Soc Series B Stat Methodol}, 53\penalty0 (1):\penalty0
  233--243, 1991.
\newblock \doi{10.1111/j.2517-6161.1991.tb01821.x}.

\bibitem[Scheipl et~al.(2015)Scheipl, Staicu, and Greven]{Scheipl.etal_2015}
F.~Scheipl, A.-M. Staicu, and S.~Greven.
\newblock {Functional Additive Mixed Models}.
\newblock \emph{J Comput Graph Stat}, 24\penalty0 (2):\penalty0 477--501, 2015.
\newblock \doi{10.1080/10618600.2014.901914}.

\bibitem[Scheipl et~al.(2016)Scheipl, Gertheiss, and Greven]{Scheipl.etal_2016}
F.~Scheipl, J.~Gertheiss, and S.~Greven.
\newblock {Generalized functional additive mixed models}.
\newblock \emph{Electron J Stat}, 10\penalty0 (1):\penalty0 1455 -- 1492, 2016.
\newblock \doi{10.1214/16-EJS1145}.

\bibitem[Wang et~al.(2016)Wang, Chiou, and Müller]{Wang.etal_2016}
J.-L. Wang, J.-M. Chiou, and H.-G. Müller.
\newblock {F}unctional {D}ata {A}nalysis.
\newblock \emph{Annu Rev Stat Appl}, 3:\penalty0 257--295, 2016.
\newblock \doi{10.1146/annurev-statistics-041715-033624}.

\bibitem[Wang et~al.(2022)Wang, Yang, Yi, Zhang, and Han]{Wang.etal_2022}
Z.~Wang, D.-H. Yang, T.-H. Yi, G.-H. Zhang, and J.-G. Han.
\newblock {Eliminating environmental and operational effects on structural
  modal frequency: A comprehensive review}.
\newblock \emph{Struct Control Health Monit}, 29\penalty0 (11):\penalty0 e3073,
  2022.
\newblock \doi{10.1002/stc.3073}.

\bibitem[Wittenberg et~al.(2024)Wittenberg, Knoth, and
  Gertheiss]{Wittenberg.etal_2024}
P.~Wittenberg, S.~Knoth, and J.~Gertheiss.
\newblock {Structural Health Monitoring with Functional Data: Two Case
  Studies}.
\newblock \emph{arXiv:stat.AP/2406.01262}, 2024.
\newblock \doi{10.48550/arXiv.2406.01262}.

\bibitem[Wood(2011)]{Wood_2011}
S.~N. Wood.
\newblock Fast stable restricted maximum likelihood and marginal likelihood
  estimation of semiparametric generalized linear models.
\newblock \emph{J R Stat Soc Series B Stat Methodol}, 73\penalty0 (1):\penalty0
  3--36, 2011.
\newblock \doi{10.1111/j.1467-9868.2010.00749.x}.

\bibitem[Wood(2017)]{Wood_2017}
S.~N. Wood.
\newblock \emph{{Generalized Additive Models: An Introduction with R}}.
\newblock CRC Press, Boca Raton, 2nd. edition, 2017.

\bibitem[Woodall(2007)]{Woodall_2007}
W.~H. Woodall.
\newblock Current research on profile monitoring.
\newblock \emph{Production}, 17\penalty0 (3):\penalty0 420–425, 2007.
\newblock \doi{10.1590/S0103-65132007000300002}.

\bibitem[Woodall et~al.(2004)Woodall, Spitzner, Montgomery, and
  Gupta]{Woodall.etal_2004}
W.~H. Woodall, D.~J. Spitzner, D.~C. Montgomery, and S.~Gupta.
\newblock Using control charts to monitor process and product quality profiles.
\newblock \emph{J Qual Technol}, 36\penalty0 (3):\penalty0 309--320, 2004.
\newblock \doi{10.1080/00224065.2004.11980276}.

\bibitem[Worden and Cross(2018)]{Worden.Cross_2018}
K.~Worden and E.~Cross.
\newblock On switching response surface models, with applications to the
  structural health monitoring of bridges.
\newblock \emph{Mech Syst Signal Process}, 98:\penalty0 139--156, 2018.
\newblock \doi{10.1016/j.ymssp.2017.04.022}.

\bibitem[Xia et~al.(2017)Xia, Zhang, Tian, and Zhang]{Xia.etal_2017}
Q.~Xia, J.~Zhang, Y.~Tian, and Y.~Zhang.
\newblock {Experimental Study of Thermal Effects on a Long-Span Suspension
  Bridge}.
\newblock \emph{J. Bridge Eng}, 22\penalty0 (7):\penalty0 04017034, 2017.
\newblock \doi{10.1061/(ASCE)BE.1943-5592.0001083}.

\bibitem[Xia et~al.(2006)Xia, Hao, Zanardo, and Deeks]{Xia.etal_2006}
Y.~Xia, H.~Hao, G.~Zanardo, and A.~Deeks.
\newblock {Long term vibration monitoring of an RC slab: Temperature and
  humidity effect}.
\newblock \emph{Eng Struct}, 28\penalty0 (3):\penalty0 441--452, 2006.
\newblock \doi{10.1016/j.engstruct.2005.09.001}.

\bibitem[Xia et~al.(2012)Xia, Chen, Weng, Ni, and Xu]{Xia.etal_2012}
Y.~Xia, B.~Chen, S.~Weng, Y.-Q. Ni, and Y.-L. Xu.
\newblock Temperature effect on vibration properties of civil structures: a
  literature review and case studies.
\newblock \emph{J Civil Struct Health Monit}, 2:\penalty0 29--46, 2012.
\newblock \doi{10.1007/s13349-011-0015-7}.

\bibitem[Yao et~al.(2005)Yao, Müller, and Wang]{Yao.etal_2005}
F.~Yao, H.-G. Müller, and J.-L. Wang.
\newblock {Functional Data Analysis for Sparse Longitudinal Data}.
\newblock \emph{J Am Stat Assoc}, 100\penalty0 (470):\penalty0 577--590, 2005.
\newblock \doi{10.1198/016214504000001745}.

\bibitem[Zhou et~al.(2011)Zhou, Serban, and Gebraeel]{Zhou.etal_2011}
R.~R. Zhou, N.~Serban, and N.~Gebraeel.
\newblock {Degradation modeling applied to residual lifetime prediction using
  functional data analysis}.
\newblock \emph{Ann Appl Stat}, 5\penalty0 (2B):\penalty0 1586--1610, 2011.
\newblock \doi{10.1214/10-AOAS448}.

\end{thebibliography}
\end{document}